\pdfoutput=1
\documentclass[lettersize,journal]{IEEEtran}
\let\appendices\relax
\usepackage{amsmath,amsfonts}
\usepackage{algorithmic}
\usepackage{algorithm}
\usepackage{array}
\usepackage[caption=false,font=normalsize,labelfont=sf,textfont=sf]{subfig}
\usepackage{textcomp}
\usepackage{stfloats}
\usepackage{url}
\usepackage{verbatim}
\usepackage{graphicx}
\usepackage{cite}
\usepackage{lipsum}
\usepackage{float}
\usepackage{rotating}
\usepackage{threeparttable}
\usepackage{booktabs}
\usepackage{multirow}
\usepackage{longtable}
\usepackage{pdflscape}
\usepackage{appendix}
\usepackage{balance}

\newcolumntype{L}[1]{>{\raggedright\let\newline\\\arraybackslash\hspace{0pt}}m{#1}}
\newcolumntype{C}[1]{>{\centering\let\newline\\\arraybackslash\hspace{0pt}}m{#1}}
\newcolumntype{R}[1]{>{\raggedleft\let\newline\\\arraybackslash\hspace{0pt}}m{#1}}

\begin{document}
\title{Tool Compensation and User Strategy during Human-Robot Teleoperation are Impacted by System Dynamics and Kinesthetic Feedback}

\author{Jacob Carducci, Jeremy D. Brown
\thanks{J. Carducci and J.D. Brown are with the Department of Mechanical Engineering, Johns Hopkins University, Baltimore, MD, 21218 USA email: jdelainebrown@jhu.edu}
\thanks{This work has been submitted to the IEEE for possible publication. Copyright may be transferred without notice, after which this version may no longer be accessible.}}

\maketitle

\begin{abstract}
Manipulating an environment remotely with a robotic teleoperator introduces novel electromechanical (EM) dynamics between the user and environment. While considerable effort has focused on minimizing these dynamics, there is limited research into understanding their impact on a user’s internal model and resulting motor control strategy. Here we investigate to what degree, if any, the dynamics and kinesthetic feedback of the teleoperator influence task behavior and tool compensation.

Our teleoperator testbed features a leader port controlled by user input via wrist pronation/supination, a follower port connected to a virtual environment rendered by a rotary motor, and three distinct transmissions in-between that can be engaged independently.  A Rigid transmission is created with a rigid rod that mechanically couples the leader and follower without scaling. A Unilateral transmission is created with two rotary motors that couple the leader and follower without force feedback to the user. A Bilateral transmission is also created with these rotary motors by providing force cues back to the user during leader-follower coupling.

N~=~30 adult participants rotated a virtual disk in a visco-elastic virtual environment through counterbalanced presentation of each transmission mode. Users tracked targets oscillating at pre-defined frequencies, randomly selected from 0.55, 0.85, 1.15, 1.55, 1.85, 2.05, and 2.35\,Hz, with an initial training frequency at 1.25\,Hz. After session completion, trajectories of the target, leader, and follower were decomposed into components of gain and phase error for all frequencies. We found that while tracking performance at the follower port was similar across transmissions, users' adjustment at the leader port differed between Rigid and EM transmissions. Users applied different pinch forces between Rigid and Unilateral transmissions, suggesting that tracking strategy does change between dynamics and feedback. However, the users' ability to compensate dynamics diminished significantly as task speed got faster and more difficult. Therefore, there are limits to pursuit tracking at the human wrist when compensating teleoperator dynamics.

\end{abstract}

\begin{IEEEkeywords}
Teleoperation, tracking, transmission, wrist, leader, follower, compensation, strategy, dynamics, sensorimotor, robotics, hand, adjustment
\end{IEEEkeywords}

\section{Introduction}
Teleoperator tools are devices operated by a user that augment input over distance, through scaling, or provide a safety barrier from the output. In other words, teleoperation extends the human capacity to sense and manipulate into remote environments \cite{Sheridan1989Telerobotics}. Originally, teleoperators involved direct mechanical connections with simple tools that allowed for basic reflection of force feedback from    the environment \cite{Goertz1949Master-SlaveManipulator}. The limited workspace and capabilities of teleoperators were later enhanced or replaced by active motor electronics, which could simulate and render kinetics imparted by the environment through the output tool to the user \cite{Niemeyer2008Telerobotics}. There have been various historical reviews of teleoperation \cite{Hokayem2006BilateralSurvey,Sheridan1989Telerobotics,Vertut1986TeleoperationsDevelopment}, which readers are encouraged to explore. Robotic teleoperation has been explored in various fields such as rehabilitation \cite{Culmer2010ASystem,Atashzar2017AControl,Rahman1993BilateralRobot}, prosthetics \cite{Welker2021TeleoperationExoskeleton}, surgery \cite{Burgner2014ASurgery,Greer2008HumanMachineStereotaxy,Wagner2007TheDissection}, medical training \cite{LOrsa2013IntroductionNeurosurgeons,Tahmasebi2008ASimulator,Shahbazi2018MultimodalTraining}, oceanic \cite{Khatib2016OceanDiscovery} and space exploration \cite{Sheridan1993SpacePrognosis,Penin2000ForceRobots}, hazardous material handling \cite{Kron2004DisposalSystem,Goertz1949Master-SlaveManipulator}, construction \cite{Zhu2021NeurobehavioralTeleoperation}, and disaster rescue \cite{Negrello2018HumanoidsScenario}.

The transfer of power and information between the user-facing leader port and the environment-facing follower port is required for single-input single-output (SISO) telemanipulators to operate. Unilateral teleoperators transmit leader action to the follower touching the environment, but no information or power is sent back. Bilateral teleoperators transmit cues of environmental events through the follower to the leader as feedback. Usually feedback in teleoperators are presented as kinesthetic haptic forces, which provide the perception of weight or resistance felt by the user from the environment \cite{Burdea1996ForceReality}. The reflection of transparent force feedback tends to improve task performance measures such as completion time, accuracy, and cognitive load \cite{Hannaford1991PerformanceTeleoperator,Massimino1994TeleoperatorFeedback,Zhu2021NeurobehavioralTeleoperation,OMalley2006SharedTraining} compared to unilateral teleoperation without feedback. Integrating multimodal feedback such as vision can further improve task performance \cite{Mitsou2006Visuo-HapticTasks}.

When the user interacts with the leader port, the human and teleoperator device create a coupled system. This coupled interaction integrates the impedance of the user’s arm, which can stabilize the system by dissipating energy from reactive, elastic biomechanics \cite{Mussa-Ivaldi1985NeuralHumans,Tsuji1995HumanPosture}. By itself, the user’s arm violates passivity from its force-generating muscle actuators and neural feedback \cite{Hogan1989ControllingInterface}. Interestingly, stabilization from introducing user-modulated arm impedance generally makes the combined system more passive than just the teleoperative tool. A contributing factor is that the grasping interface of the user can absorb energy from the telemanipulator. Relaxed grasping has a net drain on stored system energy due to dampening mechanics \cite{Dyck2013IsPassive}.

How can the human actively accomplish task goals while maintaining system stability and passivity during manipulation? The key is that the human arm is a neuromusculoskeletal system. The central nervous system (CNS) compares sensory input predicted from motor reafference to actual sensory input to build and update an internal model of interaction \cite{Tong2003Task-SpecificTransformations,Fuentes2010WhereTasks}. Using the arm as a neurological tool, the brain can learn from environments, update subjective experience, and adapt motor strategies based on kinematic goals \cite{Kawato1999InternalPlanning}. Specifically, the human CNS can internalize an compensatory inverse model of the environment to cancel out non-linear dynamics \cite{Robinson2016DynamicSystems}. The CNS can then modify impedance of the arm from muscle co-activation to restabilize the interaction, and update the inverse model from sensory mismatch to improve general robustness \cite{Franklin2003AdaptationModel,Burdet2001TheImpedance}. The user generally starts learning a specific task by modifying their motor control over time: the strategy starts usually with a high-impedance arm from co-contracting muscles, which progressively relaxes as the internal compensatory task model is formed \cite{Heald2018IncreasingLearning,Gribble2003RoleAccuracy}.

Despite advances in control and neuroscience, there has been limited research effort into understanding how specifically a person’s motor strategy during environmental manipulation, especially in a wrist-focused tracking task, is impacted by the dynamics of a mediating device in-between \cite{Nisky2014EffectsSurgeons,Markovic2018TheProsthesis}. In other words, does adding teleoperation influence different task behavior from a user? How does modifying the teleoperator transparency and mechanics impact the user’s compensation of the teleoperator? Our previous study \cite{Singhala2023TelerobotTask} suggests different dynamics do influence how telerobots are compensated and incorporated by users. However, only a singular tracking task was presented, so it did not test the limits of human compensation of teleoperator tools. Comparing task success between direct and electromechanical transmissions over a sequence of task difficulties as tracking speeds may shed light on the specific conditions and limits where compensation (and perhaps embodiment) breaks down during wrist tracking.

Leveraging a pre-existing teleoperator testbed with different transmission configurations, we have expanded our study to investigate the impact that transmission dynamics have on a user’s ability to track a target rotationally in a visual-haptic task using their wrist. Our hypotheses were that teleoperator dynamics, task difficulty, and task repetitions all play roles in tracking performance and motor control strategy. Our corresponding predictions were that changing transmission mode impacts tracking effectiveness from wrist pronation/supination differently, that grip force during tracking is impacted by changing transmissions, and that performance suffers as rotational target speed increases (to reaffirm prior literature).

\section{Methods}
\subsection{Experimental Setup}
\begin{figure}[H] 
   \centering
   \includegraphics[scale=0.4]{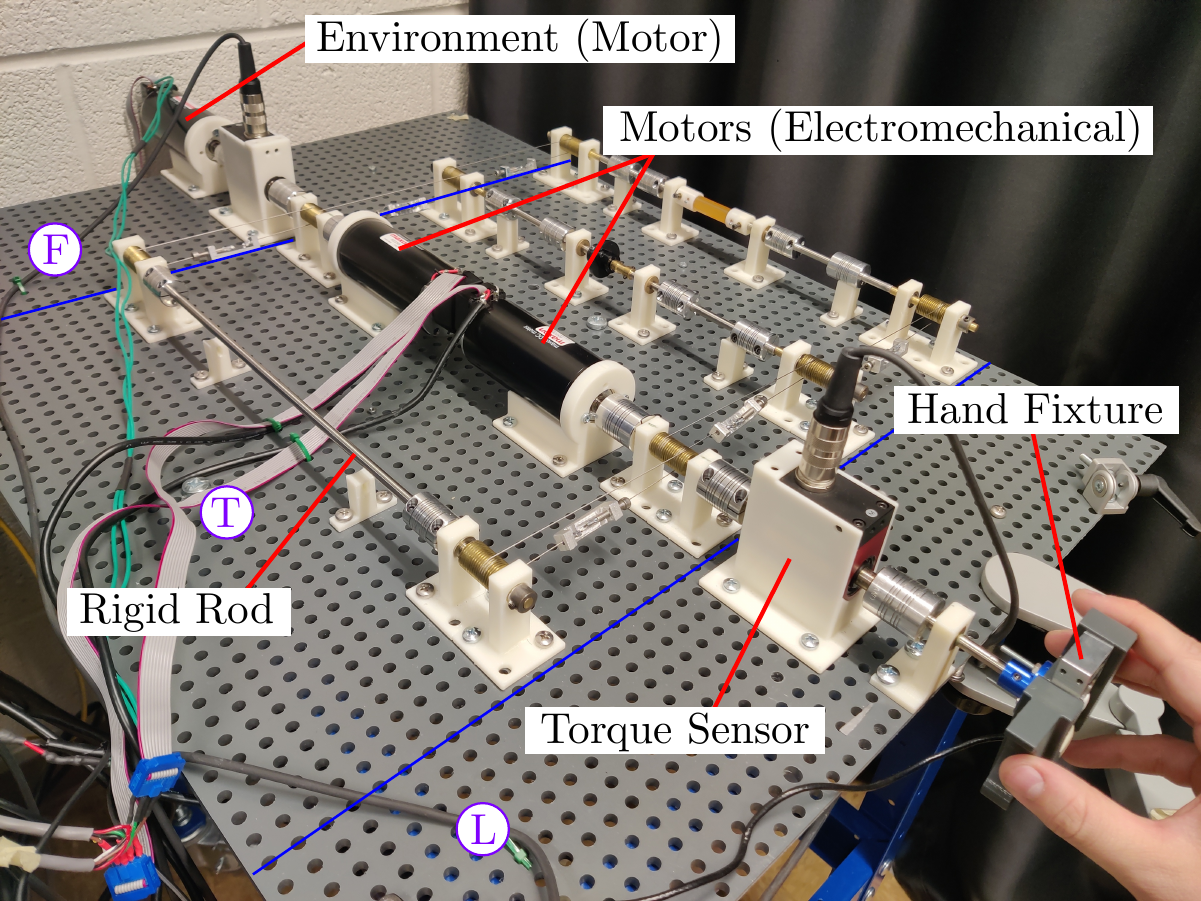}
   \caption{A general overview of the teleoperator testbed \cite{Singhala2021ADynamics}. The user interacted through a grip interface attached to the leader side (circled L), and the environment was rendered by a motor attached to the follower side (circled F). A steel rod forms the direct rigid transmission when coupled, whereas an additional motor pair in the transmission region (circled T) provides electromechanical transmission for unilateral and bilateral forcing when coupled. (The elastic and damping transmissions in the background were decoupled and ignored for this experiment.)}
   \label{fig:teleopsetup}
\end{figure}

The full experimental setup depicted in Figure~\ref{fig:teleopsetup} is detailed by Singhala and Brown \cite{Singhala2021ADynamics}. However, a summary will be provided here. The testbed has a leader side controlled by the user input and a follower side connected to the environment. The leader is equipped with a pinch interface made from 3D printed halves for the user’s hand to grab onto, and an LSP-10 beam load cell to capture gripping force. We track grip force as a simple but indirect way to measure arm activation and impedance. For the tracking experiment, the environment expressed to the follower was rendered by a Maxon RE50 motor, which can render 467\,mNm peak torque and 233\,mNm continuous torque.

Between the port interfaces, multiple transmission shafts can be engaged or disengaged to carry rotational energy between the leader and follower. To transmit energy to adjacent shafts, brass capstan pulleys are coupled to each shaft end, and adjacent pulleys are wound multiple times with 18-8 stainless steel wire. Ends of the steel wire loops are attached to aluminum U-channels with adjustable screws, which can be twisted to tighten the loops and ensure proper tension and friction on the pulleys.

Of all available transmissions, two types were utilized in the tracking experiment: Rigid and Electromechanical (EM). The Rigid transmission is a 316 stainless steel rod directly coupled to each side without any scaling of torque or position. The EM transmission incorporates two Maxon RE50 motors to inject and render torques. Each motor is equipped with a 500\,CPT HEDL encoder to track rotational position of either side. In a Unilateral configuration mode, the position of the leader motor can inform the torque of the follower motor but no feedback torque is rendered on the leader motor. In a Bilateral configuration mode, the position of the follower motor can inform the torque of the leader motor to generate feedback towards the user.

To engage with the leader and follower, a transmission has to be attached with aluminum spiral-cut couplers to corresponding capstan pulleys. The couplers can be untightened and shifted to disengage the transmission from the pulleys, disabling rotation.

All Maxon motors are current-controlled with a Quansar AMPAQ L4 amplifier, which receives commands from a Quansar QPIDe DAQ operating at 1\,kHz from Simulink 10.2 in MATLAB 2020b (Mathworks; Natick, MA, USA) with QUARC 2020 SP2 (Quanser Software; Markham, ON, CA). PD control loops are used to operate the Unilateral and Bilateral motor laws. Leader and follower shaft torques are captured with two Futek TRS600 torque sensors rated at 5\,N-m, and corresponding voltages are sent to the DAQ for handling in Simulink.

\begin{figure} 
   \centering
   \includegraphics[scale=0.45]{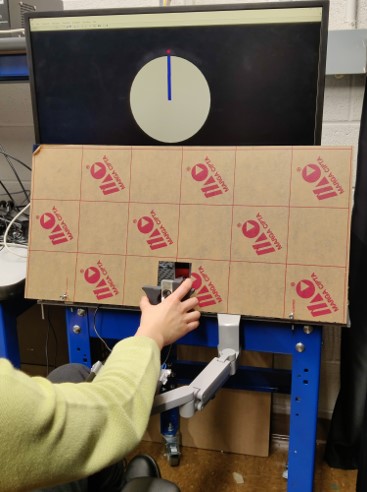}
   \caption{While the user controls the grip interface with their wrist during a tracking task, any potential sight of the testbed mechanics was obscured by a rigid opaque barrier fastened underneath the task monitor and around the input shaft. This way, visual confounds were mitigated without impacting the motion of the wrist.}
   \label{fig:teleopscreen}
\end{figure}

In addition, there were multiple peripherals used to control user vision, hearing, and posture during the tracking experiment. A high-definition monitor was used to graphically display virtual tasks to the user. An opaque acrylic screen was fastened to the front of the testbed, shown in Figure~\ref{fig:teleopscreen}, to ensure the user did not get visual rotational cues from the transmission mechanics. Noise-cancelling headphones were fitted to the user’s ears to mitigate audio cues from the rotating transmissions and any distracting noise in the testing room. A height-adjustable multi-linkage forearm rest was affixed to the front of the testbed to allow for comfortable alignment of the user’s forearm with the grip interface. A height-adjustable chair was recommended for users to allow for better comfort and forearm alignment.

\begin{figure*} 
   \centering
   \includegraphics[scale=0.45]{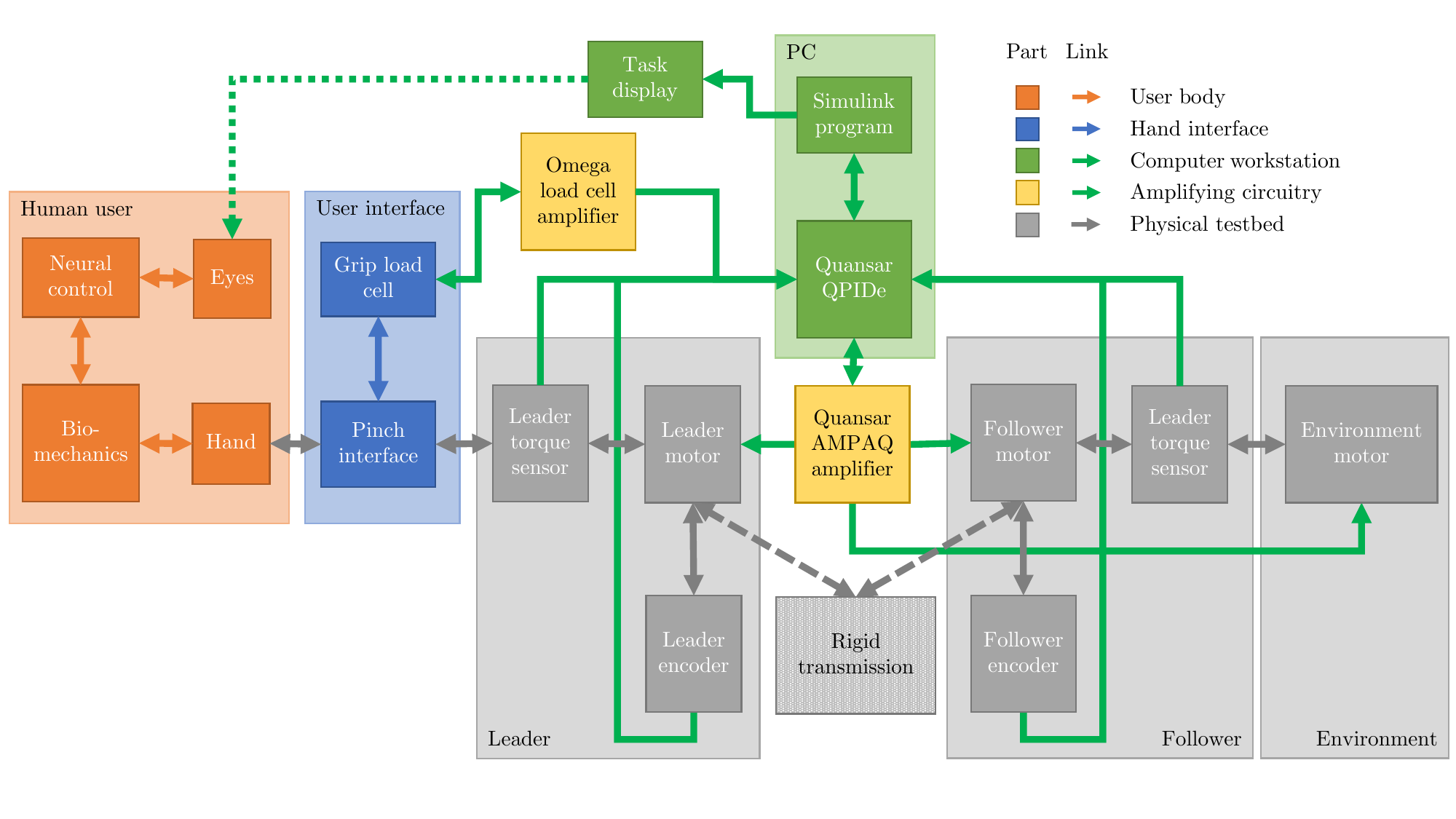}
   \caption{A graphical summary of all notable components of the human-device interaction system during the teleoperation study. Green arrows generally indicate electronic signal flows. Dashed link arrows highlight connections that can be toggled.}
   \label{fig:teleopflow}
\end{figure*}

Figure~\ref{fig:teleopflow} illustrates the direction of recorded sensor signals, physical connection, and actuating commands between different components of the experimental setup.

\subsection{Participants}
N = 30 individuals (10 F) were recruited to participate in the tracking experiment. All participants were between 18 and 65 years of age (24.8 $\pm$ 7.49), and all were affiliated with Johns Hopkins University or Johns Hopkins Hospital. Gender and age breakdowns by participant are listed in Table~\ref{tab:teleopdemograph}, along with self-reported handedness. All participants were provided written informed consent according to a protocol approved by the Johns Hopkins School of Medicine Institutional Review Board (Study\# IRB00263386). Participants were compensated at \$10/hour.

\begin{table}
  \centering
  \caption{Demographic information of the participants in the wrist teleoperator tracking experiment.}
  \begin{threeparttable}
    \begin{tabular}{lrll}
    \toprule
    PID\tnote{1}   & \multicolumn{1}{r}{Age} & \multicolumn{1}{l}{Gender} & \multicolumn{1}{l}{Handiness} \\
    \midrule
    \multicolumn{1}{l}{1} & 25    & \multicolumn{1}{l}{M} & \multicolumn{1}{l}{Ambidextrous (Right)} \\
    \multicolumn{1}{l}{2} & 27    & \multicolumn{1}{l}{M} & \multicolumn{1}{l}{Ambidextrous (Left)} \\
    \multicolumn{1}{l}{3} & 24    & \multicolumn{1}{l}{M} & \multicolumn{1}{l}{Right} \\
    \multicolumn{1}{l}{4} & 24    & \multicolumn{1}{l}{F} & \multicolumn{1}{l}{Right} \\
    \multicolumn{1}{l}{5} & 26    & \multicolumn{1}{l}{M} & \multicolumn{1}{l}{Right} \\
    \multicolumn{1}{l}{6} & 27    & \multicolumn{1}{l}{M} & \multicolumn{1}{l}{Right} \\
    \multicolumn{1}{l}{7} & 18    & \multicolumn{1}{l}{M} & \multicolumn{1}{l}{Right} \\
    \multicolumn{1}{l}{8} & 29    & \multicolumn{1}{l}{F} & \multicolumn{1}{l}{Right} \\
    \multicolumn{1}{l}{9} & 23    & \multicolumn{1}{l}{M} & \multicolumn{1}{l}{Right} \\
    \multicolumn{1}{l}{10} & 22    & \multicolumn{1}{l}{M} & \multicolumn{1}{l}{Right} \\
    \multicolumn{1}{l}{11} & 21    & \multicolumn{1}{l}{M} & \multicolumn{1}{l}{Right} \\
    \multicolumn{1}{l}{12} & 25    & \multicolumn{1}{l}{M} & \multicolumn{1}{l}{Right} \\
    \multicolumn{1}{l}{13} & 20    & \multicolumn{1}{l}{F} & \multicolumn{1}{l}{Right} \\
    \multicolumn{1}{l}{14} & 21    & \multicolumn{1}{l}{F} & \multicolumn{1}{l}{Right} \\
    \multicolumn{1}{l}{15} & 20    & \multicolumn{1}{l}{M} & \multicolumn{1}{l}{Right} \\
    \multicolumn{1}{l}{16} & 24    & \multicolumn{1}{l}{M} & \multicolumn{1}{l}{Ambidextrous (Right)} \\
    \multicolumn{1}{l}{17} & 25    & \multicolumn{1}{l}{M} & \multicolumn{1}{l}{Right} \\
    \multicolumn{1}{l}{18} & 20    & \multicolumn{1}{l}{F} & \multicolumn{1}{l}{Right} \\
    \multicolumn{1}{l}{19} & 23    & \multicolumn{1}{l}{F} & \multicolumn{1}{l}{Right} \\
    \multicolumn{1}{l}{20} & 30    & \multicolumn{1}{l}{F} & \multicolumn{1}{l}{Right} \\
    \multicolumn{1}{l}{21} & 61    & \multicolumn{1}{l}{M} & \multicolumn{1}{l}{Right} \\
    \multicolumn{1}{l}{22} & 29    & \multicolumn{1}{l}{M} & \multicolumn{1}{l}{Right} \\
    \multicolumn{1}{l}{23} & 23    & \multicolumn{1}{l}{F} & \multicolumn{1}{l}{Ambidextrous (Left)} \\
    \multicolumn{1}{l}{24} & 23    & \multicolumn{1}{l}{M} & \multicolumn{1}{l}{Right} \\
    \multicolumn{1}{l}{25} & 19    & \multicolumn{1}{l}{F} & \multicolumn{1}{l}{Right} \\
    \multicolumn{1}{l}{26} & 26    & \multicolumn{1}{l}{M} & \multicolumn{1}{l}{Ambidextrous (Right)} \\
    \multicolumn{1}{l}{27} & 19    & \multicolumn{1}{l}{M} & \multicolumn{1}{l}{Right} \\
    \multicolumn{1}{l}{28} & 23    & \multicolumn{1}{l}{M} & \multicolumn{1}{l}{Right} \\
    \multicolumn{1}{l}{29} & 24    & \multicolumn{1}{l}{M} & \multicolumn{1}{l}{Right} \\
    \multicolumn{1}{l}{30} & 23    & \multicolumn{1}{l}{F} & \multicolumn{1}{l}{Right} \\
    \midrule
    Mean  & 24.8  &       &  \\
    STD\tnote{2}   & 7.49  &       &  \\
    Median & 23.5  &       &  \\
    \bottomrule
    \end{tabular}%
    \begin{tablenotes}
    \footnotesize
    \item[1]PID: Participant Identification number; \item[2]STD: Standard deviation
    \end{tablenotes}
     \end{threeparttable}
  \label{tab:teleopdemograph}%
\end{table}%

\subsection{Study Design}
\label{subsec:design}
Upon confirmation of written informed consent, participants were seated in front of the testbed, and the seat and forearm rest were height-adjusted until the arm was comfortable and approximately co-linear with the leader shaft. Regardless of self-reported handedness, all participants used their right wrist as their preferred hand side to interact with the testbed. The user pinched onto the grip interface such that at least one finger was in contact with either 3D-printed half part, and the fingers were not contacting the load cell.

The session was divided into three blocks, where three transmission modes were counterbalanced into: Rigid, Unilateral (Electromechanical), and Bilateral (Electromechanical). In each block, rotational frequency profiles of the tracked object were pseudo-randomized into 8 subblocks, with the first block always acting as training. For this experiment, 1.25\,Hz was selected for training based on previous pilot results. Training was repeated for 2 trials, whereas non-training tracking trials were repeated for 4 trials. Design of rotational speed profiles were adapted from a paper on whole-arm reaching, but can generalize to any periodic target tracking \cite{Zimmet2020CerebellarReaching}. With a fundamental frequency of 0.05\,Hz, single-sine tracking frequencies were chosen as prime multiples of the fundamental speed. Specifically, the six frequencies of 0.55\,Hz, 1.15\,Hz, 1.55\,Hz, 1.85\,Hz, 2.05\,Hz, and 2.35\,Hz were selected for single-sine presentation. (This set of frequencies were chosen due to interesting performance differences during pilot testing of a larger frequency range.) Amplitudes of these periodic tracking signals were determined using a piecewise plateau-inverse-law function of frequency. More specifically, lower frequencies below training frequency shared the same periodic amplitude, but higher frequencies above 1.25\,Hz had periodic amplitudes proportional to the inverse of their frequency. This choice was made so participants could track at constant rotational velocity, otherwise users would have a difficult time tracking large displacements at high frequencies and arc velocities. The remaining randomized subblock was a sum-of-sines trajectory constructed from the amplitudes of all prime multiples of 0.05\,Hz up to 2.35\,Hz. The purpose of the sum-of-sines task was to set a baseline for unpredictable movement, to act as a catch trial, and to understand the individual performance of components. After all 30 trials across 8 subblocks, the test block of transmission mode concluded, the user took a survey, and took a 3-minute-minimum break before starting the next block with different transmission dynamics.

\begin{figure*} 
   \centering
   \includegraphics[scale=0.47]{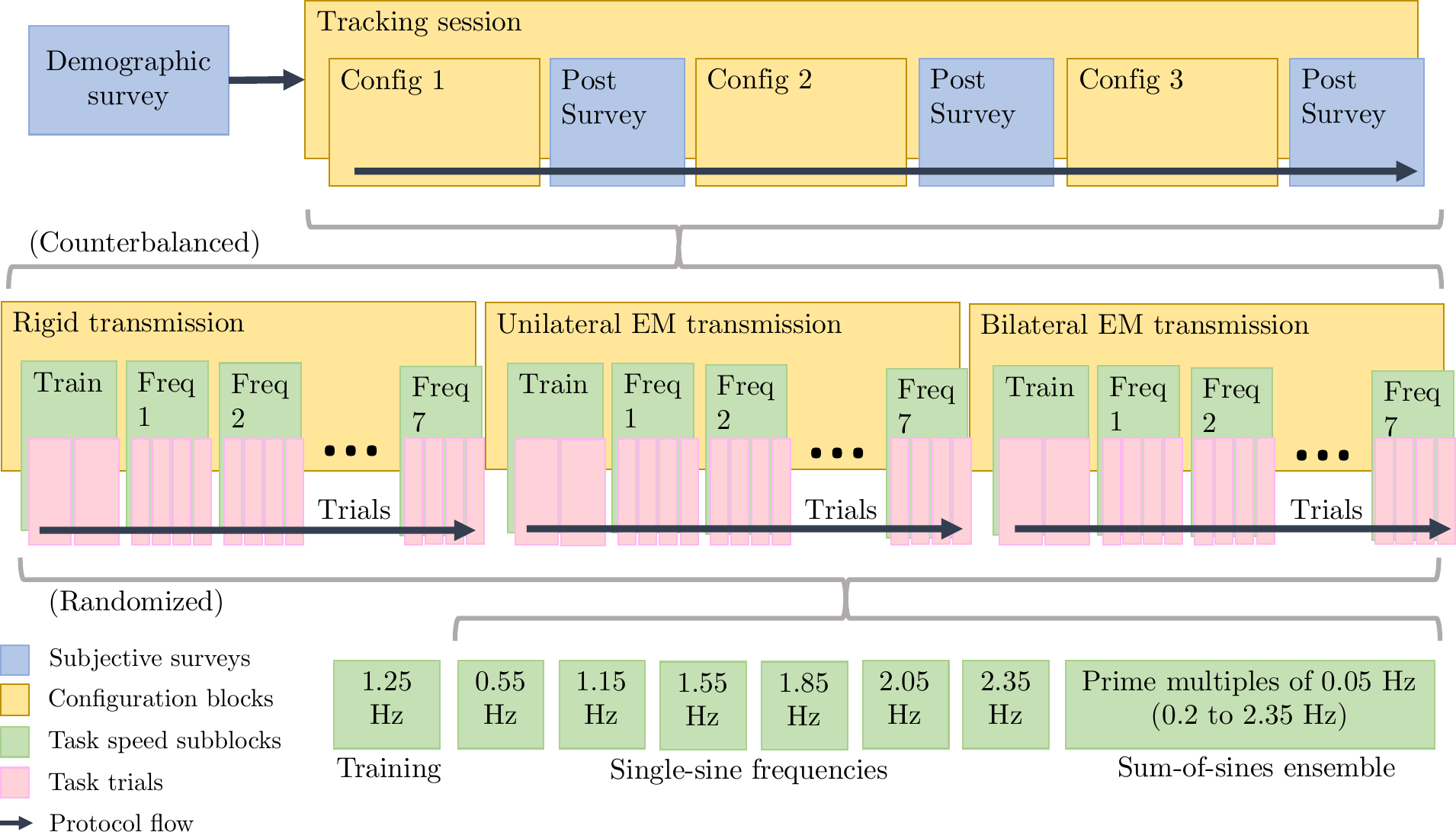}
   \caption{The protocol progression for a typical session. Note how training speeds need only two trials, whereas other task speed subblocks need four trials.}
   \label{fig:teleopdesign}
\end{figure*}

A graphical summary of a typical study session is depicted in Figure~\ref{fig:teleopdesign}.

\subsection{Tracking Task}
\begin{figure} 
   \centering
   \includegraphics[scale=0.35]{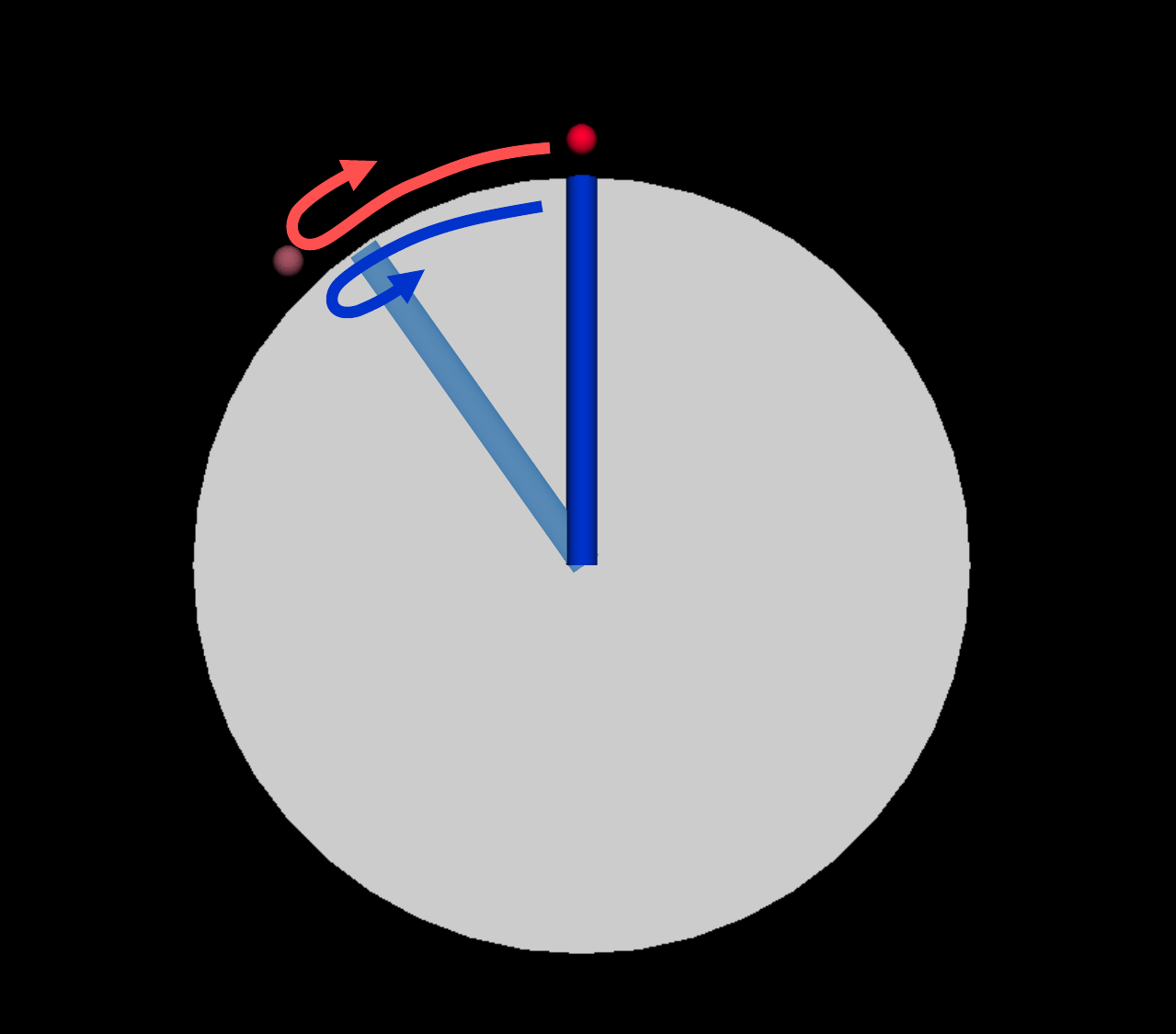}
   \caption{An example display of object-tracking at two points in time, the initial object positions in solid color and the projected positions in faded color. As the ball followed a path indicated by the red arrow, the user attempted to match the positon of the ball as indicated by the blue arrow. The viscoelastic disk is the gray circle in the background.}
   \label{fig:teleoptask}
\end{figure}

During each block, participants were told to control the pivot of a virtual rotational pointer using the rotation of the grip interface. The pointer represented the output follower position of the teleoperation testbed, not the user input directly. The pointer was affixed to a disk object, itself connected to a virtual torsional spring of 116.75\,mNm/deg and a virtual dampener of 0.67\,mNm/deg/s. When comfortable, the user was readied for the task and was prompted with an auditory cue stating “Start”. When a virtual ball object started orbiting the disk perimeter independently, the user had to align the pointer to the ball at each point in time as well as they could. This tracking setup is depicted in Figure~\ref{fig:teleoptask}. A 1-D rotational task at the wrist was specifically implemented rather than a more general tracing task because we wanted a simple introductory way to investigate the limits of human motor compensation. Despite the multiple degrees of freedom (DOF) offered by the human wrist, the tracking was restricted to 1-DOF of pronation/supination (by selecting this existing testbed) so that trajectory complexity would not place too much task demand on the user. There is the issue of ecological validity since any results from this task may not generalize to higher-order wrist motion or to general arm motion. (Verifying the generalization of any results from this study will be the subject of a future study with different equipment.)

The target ball always started at zero position above the disk, ramped up to a full oscillation centered around zero position after 5\,seconds, and continued oscillation for 20\,seconds at a frequency-determined amplitude within a range of $\pm$45\,degrees on the disk edge. The oscillation speed profile was determined by the current subblock’s frequency(s). The user repeated the tracking task for all trials for all subblocks for all blocks until session conclusion or a request for early study termination.

\subsection{Survey}
Before the first block, each participant had to self-report their age, gender, and handiness. After each block of transmission mode, participants gave subjective qualitative responses on a scale of 1-6 (Strongly Disagree to Strongly Agree) to their perceived effort and accuracy. The following questions were asked:
\begin{enumerate}
\item The task was easy to perform.
\item I was able to accurately track the moving ball.
\item My tracking accuracy improved with time.
\item My tracking accuracy was worse during tasks that seemed fast.
\item My tracking accuracy was better during tasks that seemed slow.
\item I relied on visual feedback for ball tracking.
\item I relied on haptic feedback for ball tracking.
\item I was in control of the disk/stick.
\item The motion of the disc/stick was unpredictable.
\end{enumerate}
Furthermore, participants were allowed to voluntarily share general comments at the conclusion of the session.

\subsection{Metrics and Statistical Analysis}
\begin{figure*} 
   \centering
   \includegraphics[scale=0.49]{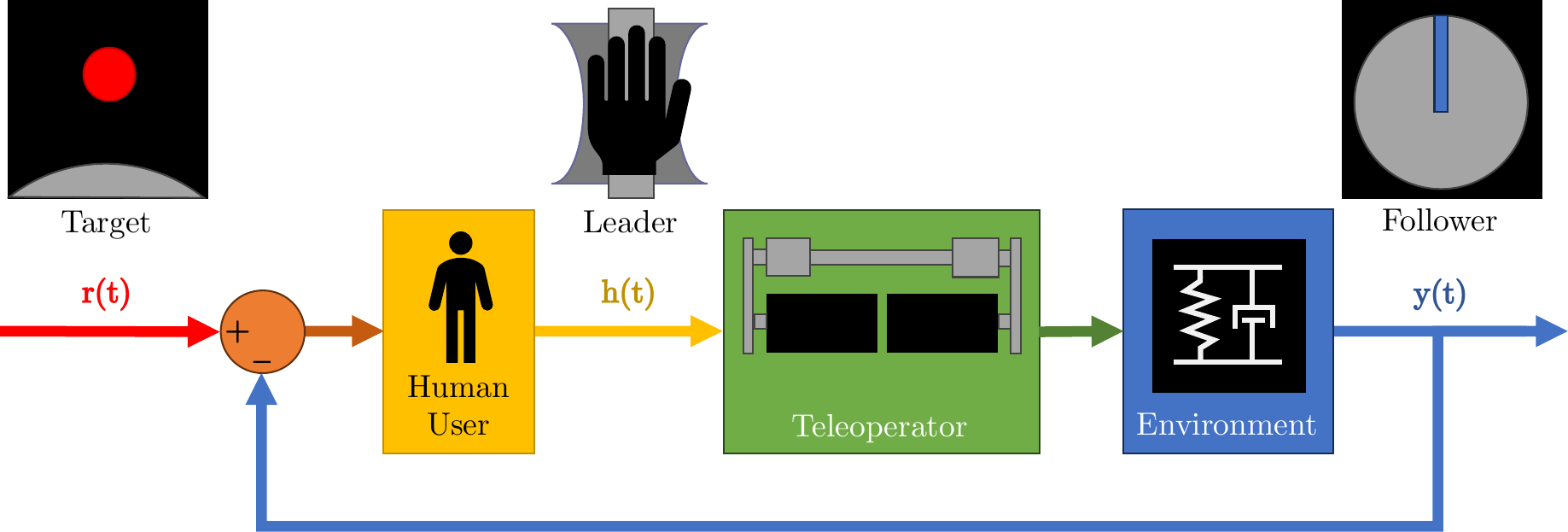}
   \caption{A simplified closed-loop control system where the human user attempts to align an output stick pointer rotation to a reference target ball rotation in real-time through a teleoperator. The difference between target and follower signals, represented as $r(t)$ and $y(t)$ respectively, is compared by the user and transformed into a hand rotation. (Inside the human is the neural controller generating control input $u(t)$ to the biomechanical plant process, but this signal has been condensed for this diagram.) The biomechanics of the hand interacting with the grip interface of the testbed leader port is represented by a leader signal $h(t)$. The leader signal is transformed through the teleoperator plant process and virtual environment plant process before outputting the follower signal $y(t)$ controlling the stick pivot. (Any inner control loops of the teleoperator process and any feedback from the leader signal have been omitted for clarity.)}
   \label{fig:teleopctrl}
\end{figure*}

\begin{figure} 
   \centering
   \includegraphics[scale=0.425]{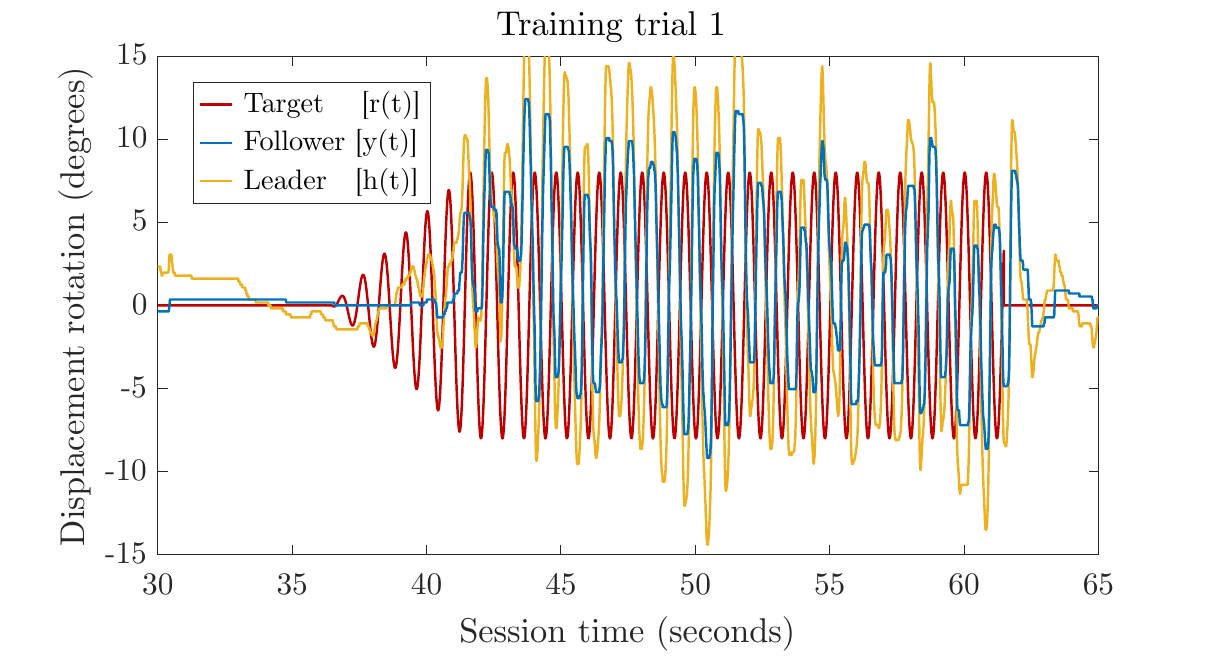}
   \caption{Rotational trajectories of direct user input on the leader side, corresponding rotation on the follower side, and the virtual target to be tracked during a given training trial. Note that the target exhibited transient ramping at trial start for 5 seconds, and maintained constant oscillation for another 20 seconds, which the user attempted to match the follower cursor to.}
   \label{fig:teleoptime}
\end{figure}

The human tracking task can be modeled as a closed loop control system as connected in Figure~\ref{fig:teleopctrl}, which illustrates the different signals of interest during wrist tracking. The ramping and oscillatory target behavior, as well as user efforts to match the trajectory, is graphed in Figure~\ref{fig:teleoptime}.

\begin{figure} 
   \centering
   \includegraphics[scale=0.55]{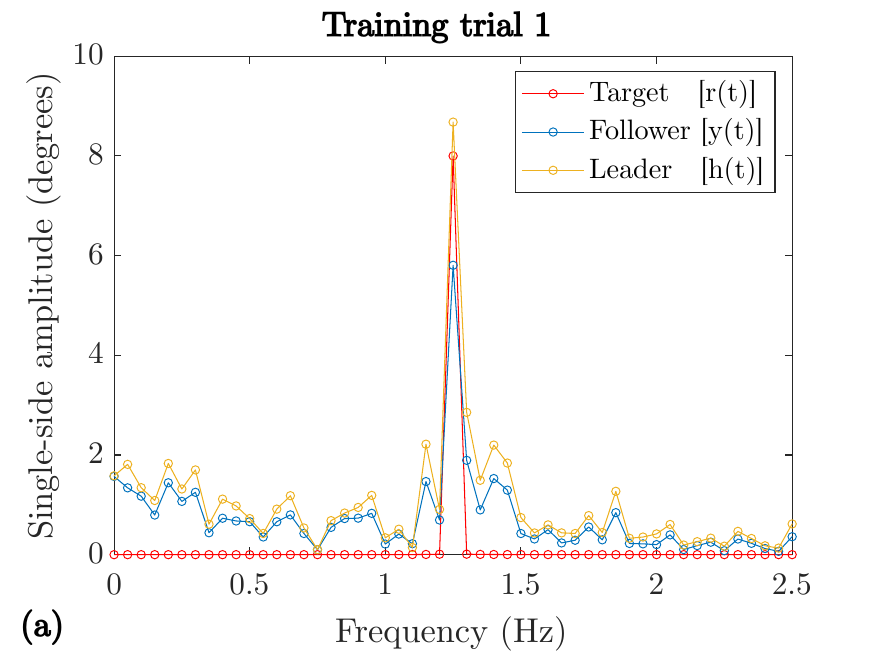}
   \includegraphics[scale=0.55]{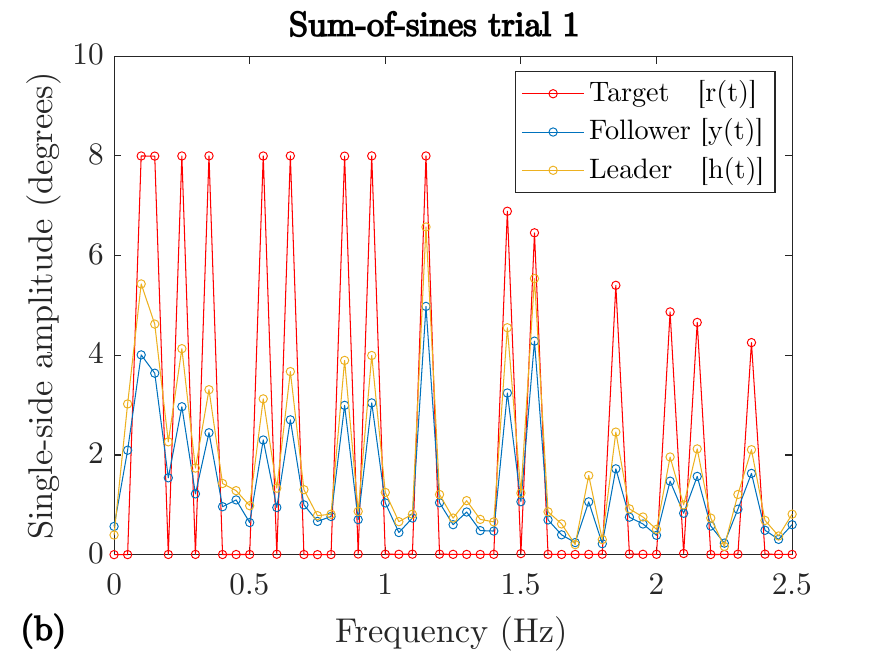}
   \caption{$\mathcal{}$ Spectral decompositions of time-series rotational trajectories from the virtual target $r(t) \xrightarrow{\mathcal{F}} R(f)$, user-controlled leader shaft $h(t) \xrightarrow{\mathcal{F}} H(f)$, and environment-connected follower shaft $y(t) \xrightarrow{\mathcal{F}} Y(f)$ from an example trial during (a) a training task and (b) a sum-of-sines task. Markers indicate magnitudes at prime multiples of the fundamental frequency (0.05 Hz). (Note the ball trajectory magnitudes obeying the piecewise scaling law during the sum-of-sines task.)}
   
   \label{fig:teleopFFT}
\end{figure}

Tracking interpretation is also adapted from the aforementioned whole-arm reaching study \cite{Zimmet2020CerebellarReaching} and an earlier animal tracking study \cite{Roth2011StimulusVirescens/i}. Each vector of time-series rotational positions from the leader encoder $h(t)$, follower encoder $y(t)$, and virtual ball $r(t)$ was converted into a single-sided spectrum of frequency components via a fast Fourier transform (FFT) implementation of the discrete Fourier transform (DFT) in MATLAB 2021a (Mathworks; Natick, MA, USA). (For example, the FFT of signal $x(t)$ is $X(f) = \mathcal{F}\{x(t)\}(f)$, where $\mathcal{F}$ is the FFT operator and $f$ is frequency in Hertz.) Complex-valued FFTs for the leader signal, follower signal, and virtual target are $H(f)$, $Y(f)$, and $R(f)$ respectively. Example sequences of single-sided amplitude profiles for each frequency of interest are shown in Figure~\ref{fig:teleopFFT}.  Note the piecewise amplitude profile of the target FFT $R(f)$ in Figure~\ref{fig:teleopFFT}b, which is the same function of bounded rotational velocity discussed in Section~\ref{subsec:design}. After frequency-domain conversion, complex-valued phasors of all three signals at task-specific frequencies were extracted so that profiles of gain (phasor magnitude) and phase (phasor angle) could be compared. The leader-target transfer function ratio is defined as $TF_{L}(f) = H(f) / R(f)$. The follower-target transfer function ratio is defined as $TF_{Y}(f) = Y(f) / R(f)$. These ratios of relative gains and phases act as tracking error metrics to determine how well the teleoperator port shaft rotations (leader or follower) matched the virtual target rotation. A perfect tracking gain ratio is 1, with higher numbers meaning the port shaft is over-rotating past the target ball, whereas lower ratios indicating the shaft is not rotating enough to the target. A perfect tracking phase ratio is 0, with lower numbers meaning the port shaft is lagging behind the target and higher ratios indicating the shaft is leading ahead of the target temporally. Grip forces for the pinch interface sensor were reported in kilograms-force, averaged at task frequencies across participants and clustered by transmission dynamics.

Linear mixed-effects models were fitted to performance measures of gain ratios, phase ratios, and grip forces, and these measures were compared pairwise using general linear hypothesis testing to reveal two-sided significant differences between transmission modes, target frequencies, trial repetitions, and covariates. When needed, tests for normality were conducted. Due to multiple comparisons, post-hoc Tukey corrections needed to be applied. For analysis, we used RStudio 2022.02.3 Build 492 (Posit Software; Boston, MA) with the lme4, lmeTest, and multcomp packages installed.

\section{Results}
While all performance measure samples passed Shapiro tests for normality, visual observation of QQ normality plots reveal heavy tailing at extreme values. Therefore, we conducted statistical analysis with nonparametric testing.

\begin{figure}[H] 
   \centering
   \includegraphics[scale=0.4]{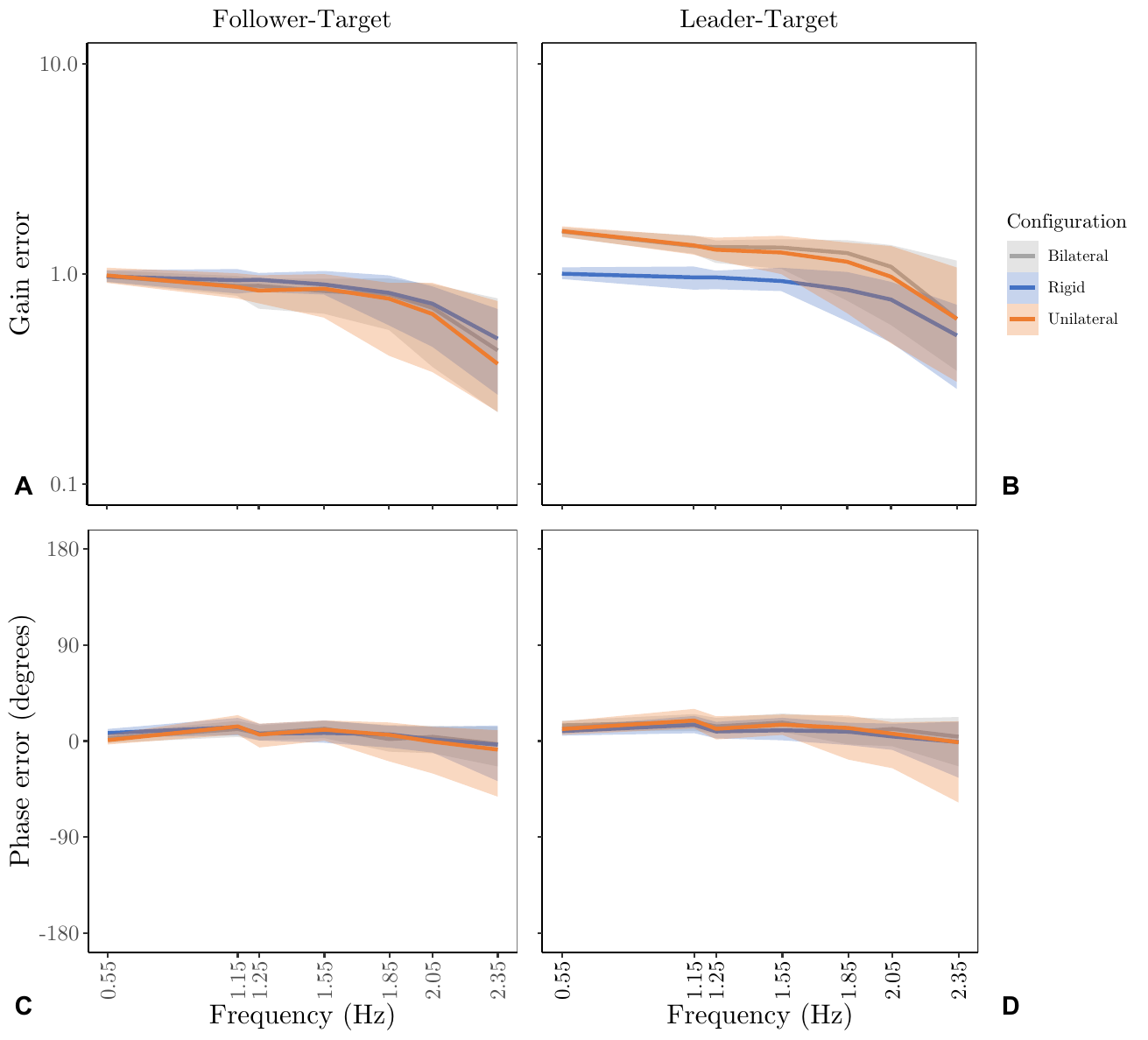}
   \caption{Ribbon-style Bode plots of median (solid line) and interquartile range (shaded area) across all single-sine task frequencies for (A) follower-target gain error, (B) leader-target gain error, (C) follower-target phase error, and (D) leader-target phase error. Error metrics were aggregated across participants and trials, and transmission modes are indicated by color.}
   \label{fig:teleoperrorresp}
\end{figure}

A graphical summary of frequency-domain error metrics from each transfer function response is shown in Figure~\ref{fig:teleoperrorresp}.

\subsection{Follower Tracking Performance is Similar Across Transmission Modes}
\begin{figure} 
   \centering
   \includegraphics[scale=0.3]{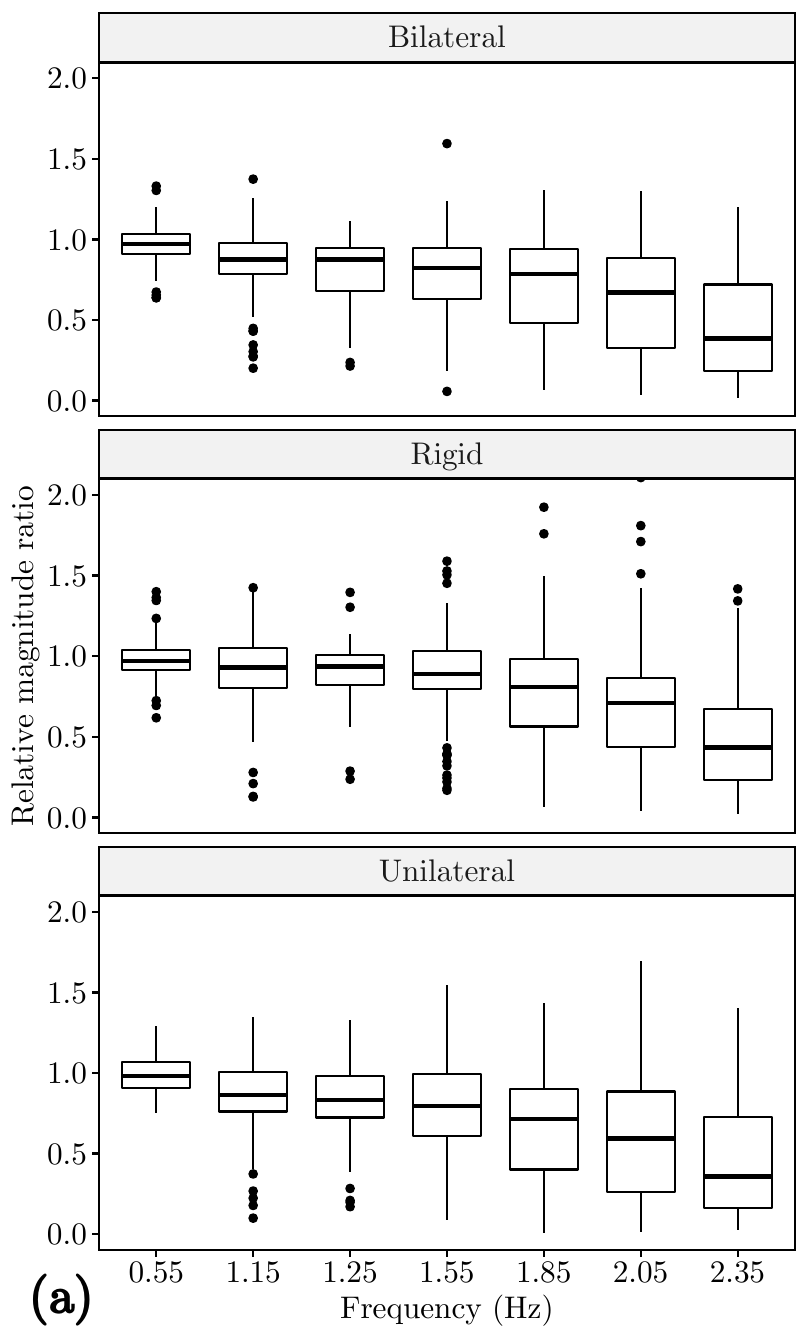}
   \includegraphics[scale=0.3]{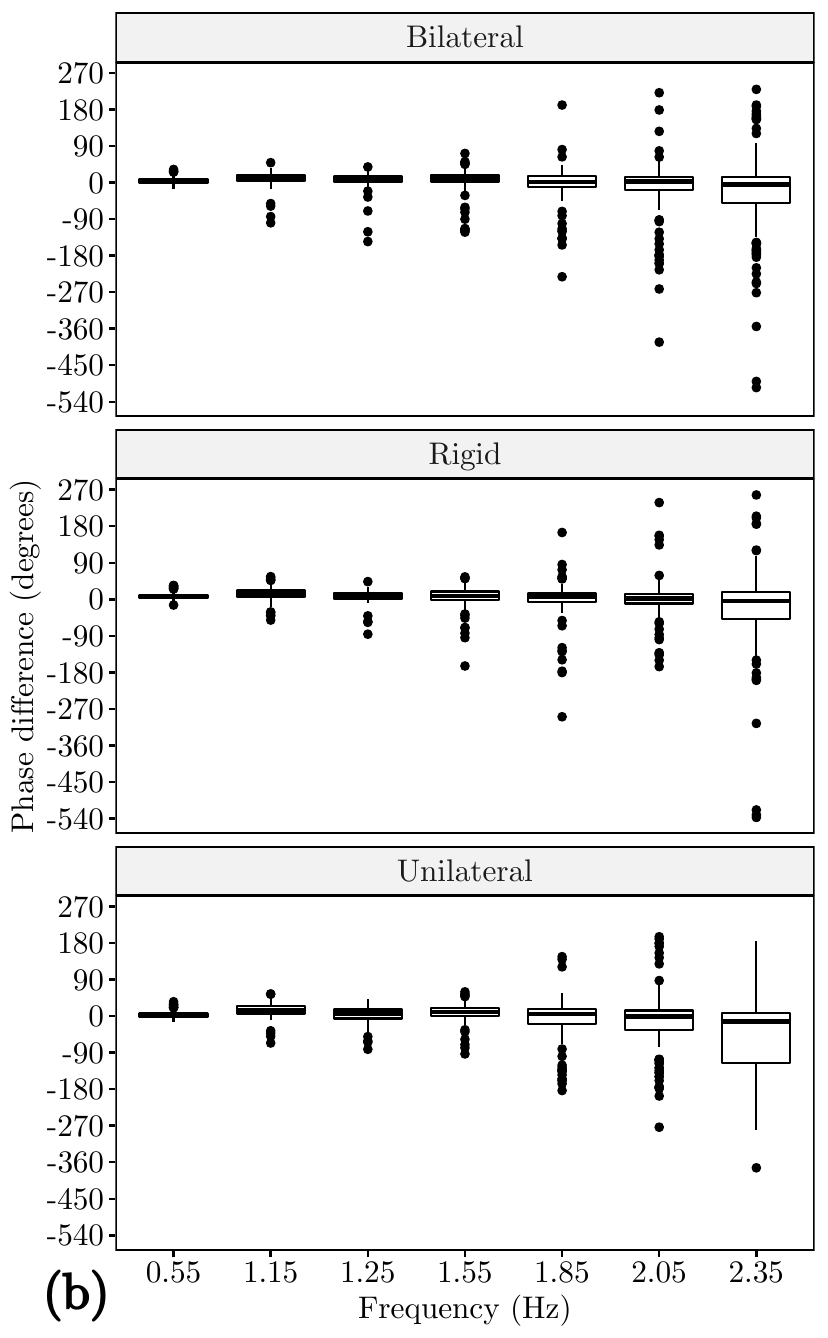}
   \caption{Modified Bode plots of the follower-target frequency response expressing relative (a) gain error and (b) phase error for all three transmission configurations by facet. Error ratios are aggregated across participants and trials. (Unlike traditional Bode plots with a continuous ``true'' scale of frequency, these plots show responses only at key discrete frequencies (axis values are not evenly spaced) for easier comparison.)}
   \label{fig:teleopFollowSiS}
\end{figure}

Between all pairwise comparisons between the three configuration modes of transmission, there were no statistically significant differences (p~$>$~0.05, listed in Tables~\ref{tab:stkabscfg} and \ref{tab:stkphscfg} of Appendix~A) in relative gain or phase between the follower and target trajectories. Supporting this, the frequency responses of the follower-target transfer function in categorical Bode gain and phase plots (shown in Figure~\ref{fig:teleopFollowSiS}) appear similar in trend across configuration modes.

\subsection{Leader Tracking Adjustment Differs Across Transmissions}
\begin{figure} 
   \centering
   \includegraphics[scale=0.3]{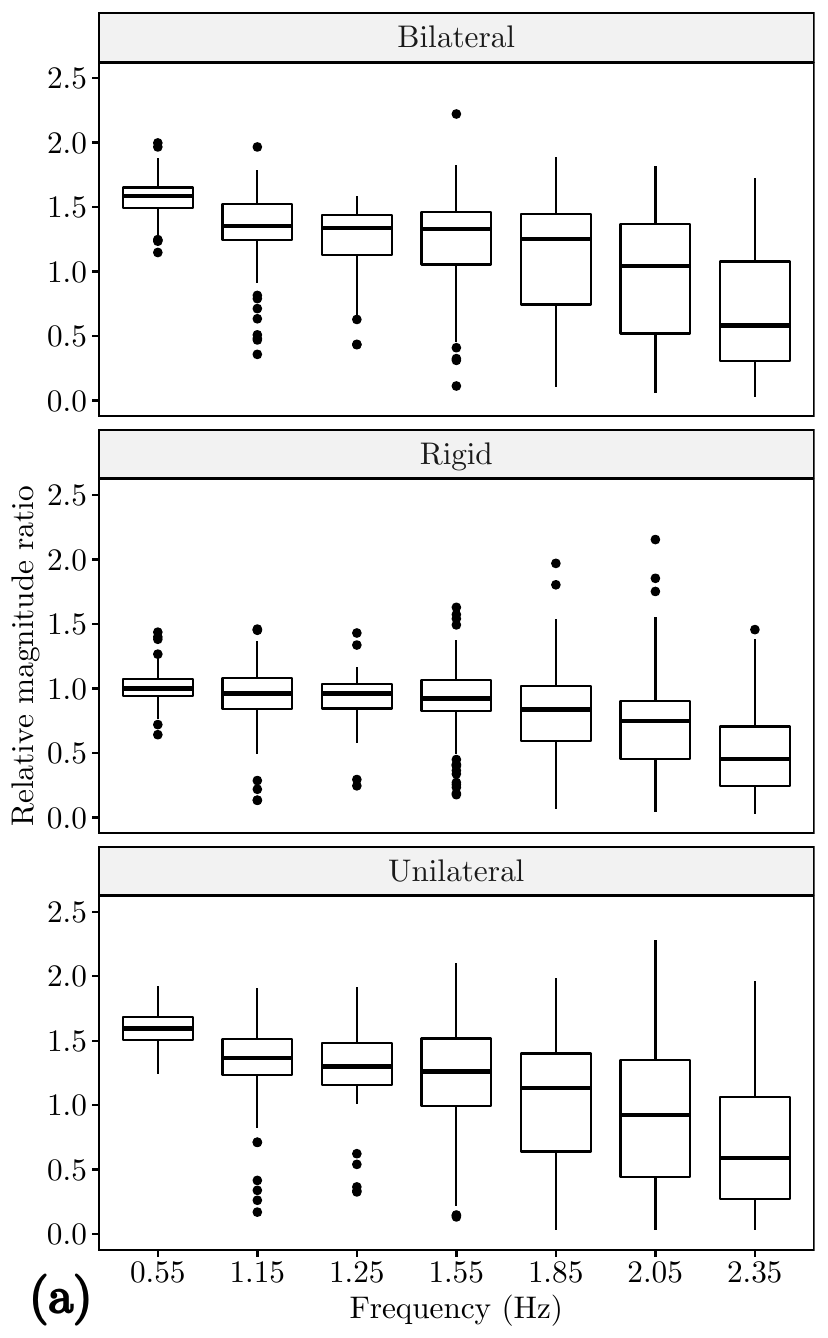}
   \includegraphics[scale=0.3]{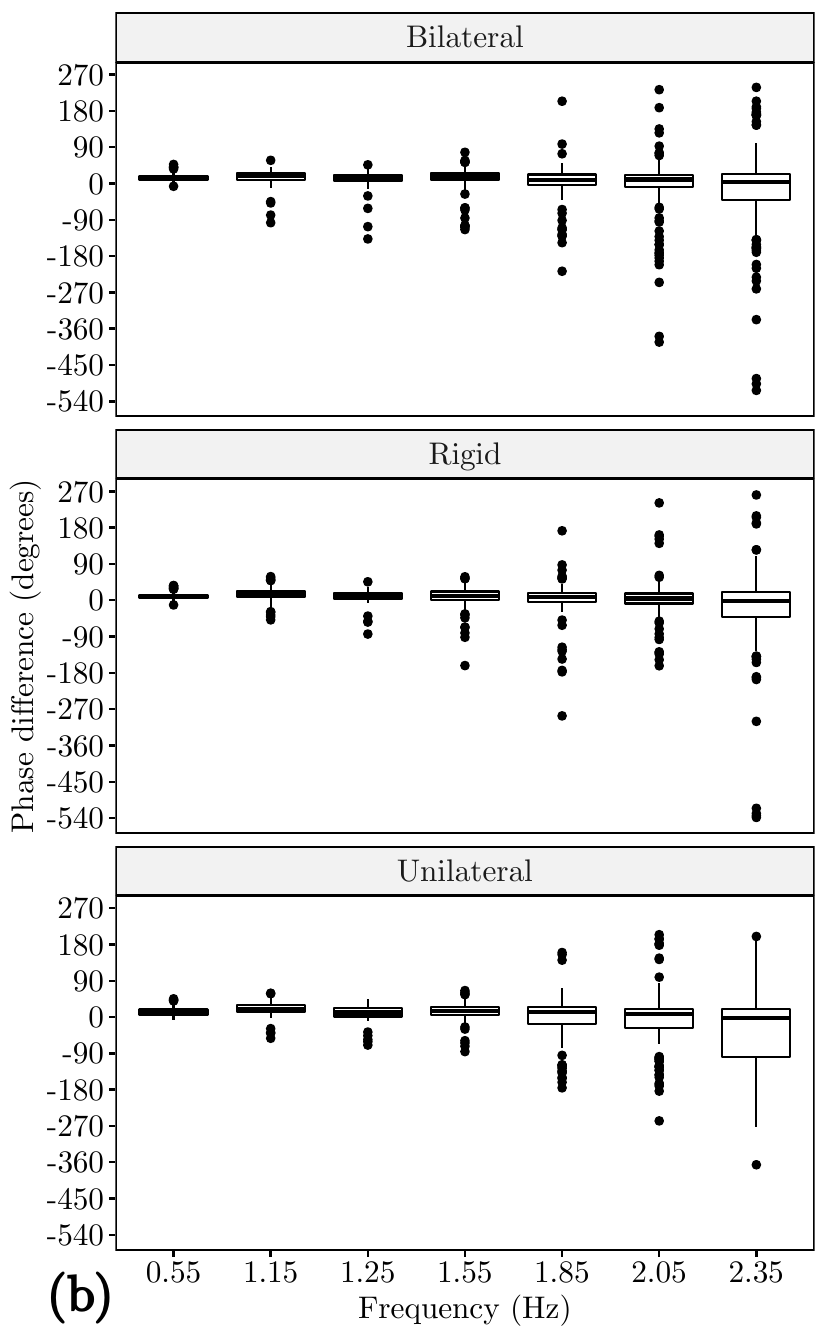}
   \caption{Modified Bode plots of the leader-target frequency response expressing relative (a) gain error and (b) phase error for all three transmission configurations by facet. Error ratios are aggregated across participants and trials.}
   \label{fig:teleopLeaderSiS}
\end{figure}

Unlike the follower-target transfer function, the leader-target transfer function did exhibit statistically significant differences in certain pairwise configuration comparisons, listed in Table~\ref{tab:ldrabscfg}. Specifically, relative gain was different between Rigid and Bilateral transmission modes (p~=~$ 1.00 \times 10^{-5}$) and between Rigid and Unilateral modes (p~=~$ 1.00 \times 10^{-5}$). The frequency responses of gain error (Figure~\ref{fig:teleopLeaderSiS}a) started higher at lower frequencies in Bilateral and Unilateral modes compared to Rigid, and descended quicker compared to Rigid.  However, no statistically significant differences were found in relative phase error (p~$>$~0.05, see Table~\ref{tab:ldrphscfg} of Appendix~B). Phase response trends between configuration modes in Figure~\ref{fig:teleopLeaderSiS}b also appear graphically similar in trend shape, suggesting approximate similarity in temporal tracking.

\begin{table}
  \centering
  \caption{P-values from pairwise comparisons of tracking gain error at leader port between configurations.}
    \begin{tabular}{lrr}
    \toprule
    Cfg & \multicolumn{1}{r}{Rigid} & \multicolumn{1}{r}{Unilateral} \\
    \midrule
    Bilateral & $ 1.00 \times 10^{-5}$ & 0.909 \\
    Rigid     &        & $ 1.00 \times 10^{-5}$\\
    \bottomrule
    \end{tabular}%
  \label{tab:ldrabscfg}%
\end{table}%

\subsection{Task Speed, Repetition and Transmission Influence Grip Force}
\begin{figure} 
   \centering
   \includegraphics[scale=0.25]{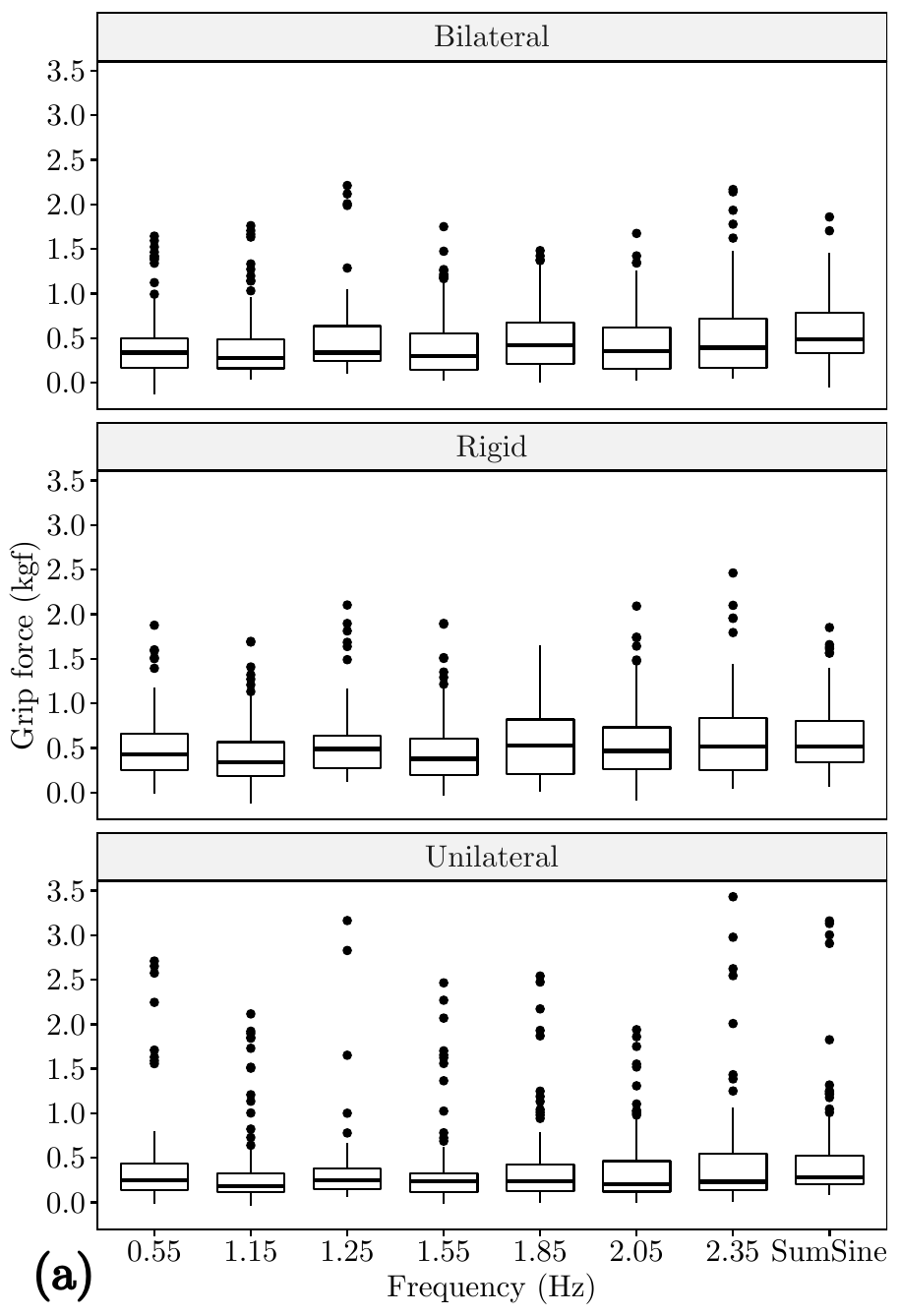}
   \includegraphics[scale=0.25]{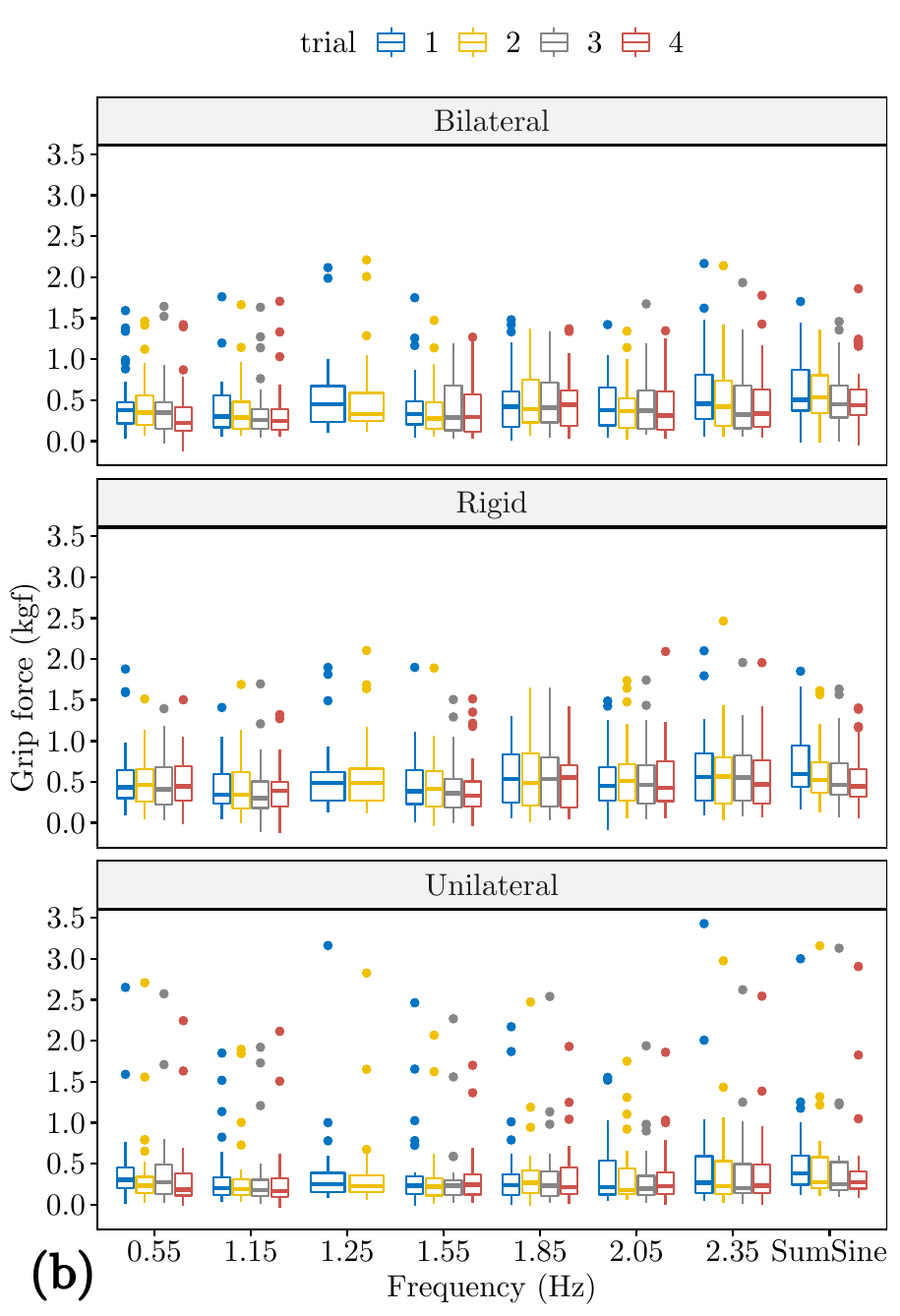}
   \caption{Modified Bode plots showing the frequency response of user pinch force for all three transmission configurations by facet. Grip forces in (a) are aggregated across participants and trials, while grip forces across participants are clustered by task repetition count in (b) with separate colors.}
   \label{fig:teleopgrip}
\end{figure}

There was a significant pairwise differences in pinch force between Rigid and Unilateral transmissions (p~=~0.007, see Table~\ref{tab:gripcfg}), despite the lack of statistically significance in corresponding fixed effects. In Figure~\ref{fig:teleopgrip}a, grip force distributions across all frequencies in Unilateral mode appear to have slighter lower interquartile ranges than the Rigid mode, even with more upper-range outliers.

\begin{table}
  \centering
  \caption{P-values from pairwise comparisons of grip force between configurations.}
    \begin{tabular}{lrr}
    \toprule
    Cfg  & \multicolumn{1}{r}{Rigid} & \multicolumn{1}{r}{Unilateral} \\
    \midrule
    Bilateral & 0.385 & 0.202 \\
    Rigid     &       & 0.007 \\
    \bottomrule
    \end{tabular}%
  \label{tab:gripcfg}%
\end{table}%

However, pinch grasps from the human user were significantly different between higher and lower single-sine frequencies (Table~\ref{tab:gripfreq}). A key observation was four significant differences in grip force between 2.35\,Hz and lower frequencies (p~=~0.001). Another important finding was significantly different grip force between the sum-of-sine task and five other single-sine tasks (p~=~0.001).

\begin{table}
  \centering
  \caption{P-values from pairwise comparisons of grip force between frequencies.}
    \begin{tabular}{lrrrrrrr}
    \toprule
    \multicolumn{1}{L{2em}}{Freq (Hz)} & 1.15  & 1.25  & 1.55  & 1.85  & 2.05  & 2.35  & \multicolumn{1}{R{2em}}{Sum Sine} \\
    \midrule
    0.55  & 0.002 & 0.054 & 0.253 & 0.147 & 1.000 & 0.001 & 0.001 \\
    1.15  &       & 0.001 & 0.779 & 0.001 & 0.001 & 0.001 & 0.001 \\
    1.25  &       &       & 0.001 & 0.989 & 0.083 & 0.871 & 0.075 \\
    1.55  &       &       &       & 0.001 & 0.168 & 0.001 & 0.001 \\
    1.85  &       &       &       &       & 0.223 & 0.113 & 0.001 \\
    2.05  &       &       &       &       &       & 0.001 & 0.001 \\
    2.35  &       &       &       &       &       &       & 0.577 \\
    \bottomrule
    \end{tabular}%
  \label{tab:gripfreq}%
\end{table}%

Furthermore, pinch grasp became significantly different by the third and fourth trial repetitions compared to the first tracking trial (p~=~0.001, see Table~\ref{tab:griptrial}). In Figure~\ref{fig:teleopgrip}b for example, the decreasing grip force distributions was apparent in most task frequencies of the Rigid transmission, with trial 1 distributions notably higher than trial 4.

\begin{table}
  \centering
  \caption{P-values from pairwise comparisons of grip force between trials.}
    \begin{tabular}{lrrr}
    \toprule
    \multicolumn{1}{l}{Trial} & 2     & 3     & 4 \\
    \midrule
    1     & 0.085 & 0.001 & 0.001 \\
    2     &       & 0.227 & 0.014 \\
    3     &       &       & 0.696 \\
    \bottomrule
    \end{tabular}%
  \label{tab:griptrial}%
\end{table}%

\subsection{Task Speed and Repetition Impacts Leader and Follower Tracking}
Unsurprisingly, all frequency responses of relative gain and phase error for both follower-target and leader-target transfer functions exhibited at least one significant difference between task speeds, shown in Tables~\ref{tab:stkabsfreq} through \ref{tab:ldrphsfreq}. A notable example is that both teleoperator ports had different error ratios between 2.35\,Hz and 0.55\,Hz for both components (p~=~0.001 in all four pairwise tables), implying performance was indeed generally different at the highest speed compared to the lowest speed.

\begin{table}
  \centering
  \caption{P-values from pairwise comparisons of tracking gain error at follower port between frequenices.}
    \begin{tabular}{lrrrrrr}
    \toprule
    \multicolumn{1}{L{2em}}{Freq (Hz)} & 1.15  & 1.25  & 1.55  & 1.85  & 2.05  & 2.35 \\
    \midrule
    0.55  & 0.001 & 0.001 & 0.001 & 0.001 & 0.001 & 0.001 \\
    1.15  &       & 0.537 & 0.011 & 0.001 & 0.001 & 0.001 \\
    1.25  &       &       & 0.957 & 0.001 & 0.001 & 0.001 \\
    1.55  &       &       &       & 0.001 & 0.001 & 0.001 \\
    1.85  &       &       &       &       & 0.001 & 0.001 \\
    2.05  &       &       &       &       &       & 0.001 \\
    \bottomrule
    \end{tabular}%
  \label{tab:stkabsfreq}%
\end{table}%

\begin{table}
  \centering
  \caption{P-values from pairwise comparisons of tracking gain error at leader port between frequencies.}
    \begin{tabular}{lrrrrrr}
    \toprule
    \multicolumn{1}{L{2em}}{Freq (Hz)} & 1.15  & 1.25  & 1.55  & 1.85  & 2.05  & 2.35 \\
    \midrule
    0.55  & 0.001 & 0.001 & 0.001 & 0.001 & 0.001 & 0.001 \\
    1.15  &       & 0.165 & 0.002 & 0.001 & 0.001 & 0.001 \\
    1.25  &       &       & 0.993 & 0.001 & 0.001 & 0.001 \\
    1.55  &       &       &       & 0.001 & 0.001 & 0.001 \\
    1.85  &       &       &       &       & 0.001 & 0.001 \\
    2.05  &       &       &       &       &       & 0.001 \\
    \bottomrule
    \end{tabular}%
  \label{tab:ldrabsfreq}%
\end{table}%

\begin{table}
  \centering
  \caption{P-values from pairwise comparisons of tracking phase error at follower port between frequencies.}
    \begin{tabular}{lrrrrrr}
    \toprule
    \multicolumn{1}{L{2em}}{Freq (Hz)} & 1.15  & 1.25  & 1.55  & 1.85  & 2.05  & 2.35 \\
    \midrule
    0.55  & 0.346 & 0.999 & 0.999 & 0.043 & 0.001 & 0.001 \\
    1.15  &       & 0.314 & 0.564 & 0.001 & 0.001 & 0.001 \\
    1.25  &       &       & 0.992 & 0.409 & 0.030 & 0.001 \\
    1.55  &       &       &       & 0.015 & 0.001 & 0.001 \\
    1.85  &       &       &       &       & 0.819 & 0.001 \\
    2.05  &       &       &       &       &       & 0.001 \\
    \bottomrule
    \end{tabular}%
  \label{tab:stkphsfreq}%
\end{table}%

\begin{table}
  \centering
  \caption{P-values from pairwise comparisons of tracking phase error at leader port between frequencies.}
    \begin{tabular}{lrrrrrr}
    \toprule
    \multicolumn{1}{L{2em}}{Freq (Hz)} & 1.15  & 1.25  & 1.55  & 1.85  & 2.05  & 2.35 \\
    \midrule
    0.55  & 0.940 & 0.928 & 0.996 & 0.005 & 0.001 & 0.001 \\
    1.15  &       & 0.436 & 0.618 & 0.001 & 0.001 & 0.001 \\
    1.25  &       &       & 0.997 & 0.505 & 0.050 & 0.001 \\
    1.55  &       &       &       & 0.044 & 0.001 & 0.001 \\
    1.85  &       &       &       &       & 0.837 & 0.001 \\
    2.05  &       &       &       &       &       & 0.001 \\
    \bottomrule
    \end{tabular}%
  \label{tab:ldrphsfreq}%
\end{table}%

\begin{figure} 
   \centering
   \includegraphics[scale=0.3]{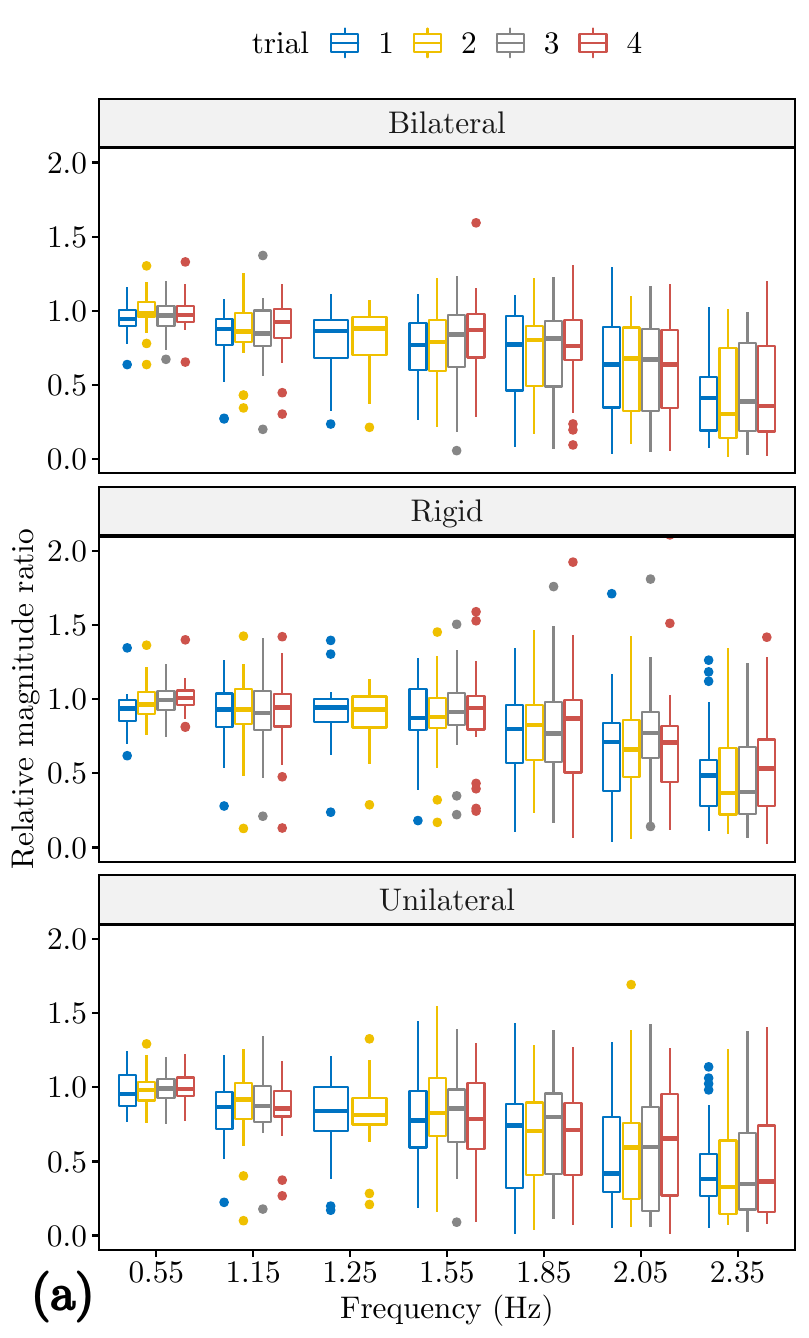}
   \includegraphics[scale=0.3]{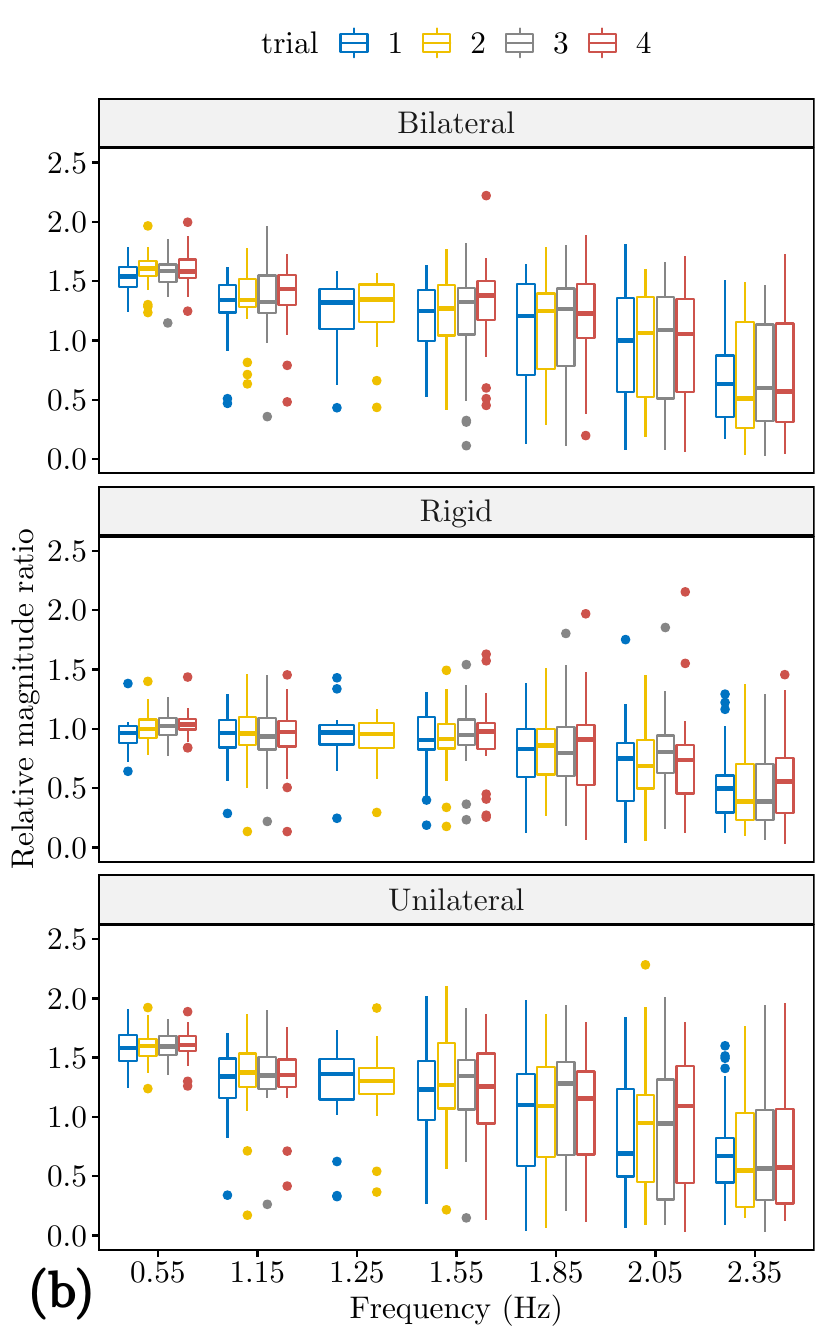}
   \caption{Modified Bode plots of the frequency responses expressing relative gain error for (a) follower-target transfer function and (b) leader-target transfer function. Responses for all three transmission configurations are indicated by facet. Error ratios are aggregated across participants, but clustered by repeated task count into separate colors.}
   \label{fig:teleopSiSmagtr}
\end{figure}

Additionally, gain responses for both transfer functions at the fourth trial differed from the first, with p~=~0.007 for follower-target (Table~\ref{tab:stkabstrial}) and p~=~0.010 for leader-target (Table~\ref{tab:ldrabstrial}). These gain error differences are most visually evident as an increase in performance in the Unilateral 2.05\,Hz condition shown in Figure~\ref{fig:teleopSiSmagtr}, where trial 4 had better tracking gain performance than trial 1. However, no statistical significance was found between trials for phase responses (p~$>$~0.05, see Tables~\ref{tab:stkphstrial} and \ref{tab:ldrphstrial} of Appendix~C).

\begin{table}
  \centering
  \caption{P-values from pairwise comparisons of tracking gain error at follower port between trials.}
    \begin{tabular}{lrrr}
    \toprule
    \multicolumn{1}{l}{Trial} & 2     & 3     & 4 \\
    \midrule
    1     & 0.166 & 0.047 & 0.007 \\
    2     &       & 0.913 & 0.582 \\
    3     &       &       & 0.931 \\
    \bottomrule
    \end{tabular}%
  \label{tab:stkabstrial}%
\end{table}%

\begin{table}
  \centering
  \caption{P-values from pairwise comparisons of tracking gain error at leader port between trials.}
    \begin{tabular}{lrrr}
    \toprule
    \multicolumn{1}{l}{Trial} & 2     & 3     & 4 \\
    \midrule
    1     & 0.223 & 0.072 & 0.010 \\
    2     &       & 0.924 & 0.551 \\
    3     &       &       & 0.904 \\
    \bottomrule
    \end{tabular}%
  \label{tab:ldrabstrial}%
\end{table}%

\subsection{Interaction Effects Occur in Grip and Tracking Response between Frequency and Transmission}
From observing the mixed-effects coefficients of our fitted linear models with interaction terms (i.e.,~frequency $\times$ configuration), we observed some statistically significant findings. For grip force, the covariate difference between Bilateral mode at 0.55\,Hz and Unilateral mode at 1.25\,Hz was significant (p~=~0.015). For follower-target gain error, the covariate difference between Bilateral at 0.55\,Hz and Rigid at 1.55\,Hz was significant (p~=~0.002). For leader-target gain error, the difference between Bilateral at 0.55 Hz and Rigid at 1.15\,Hz was significant (p~=~0.021). Finally for leader-target phase error, the difference between Bilateral at 0.55\,Hz and Rigid at 2.05\,Hz was significant (p~=~0.041).

\subsection{Users' Subjective Experience Did Not Change Much Between Transmissions}
\begin{figure*} 
   \centering
   \includegraphics[scale=0.53]{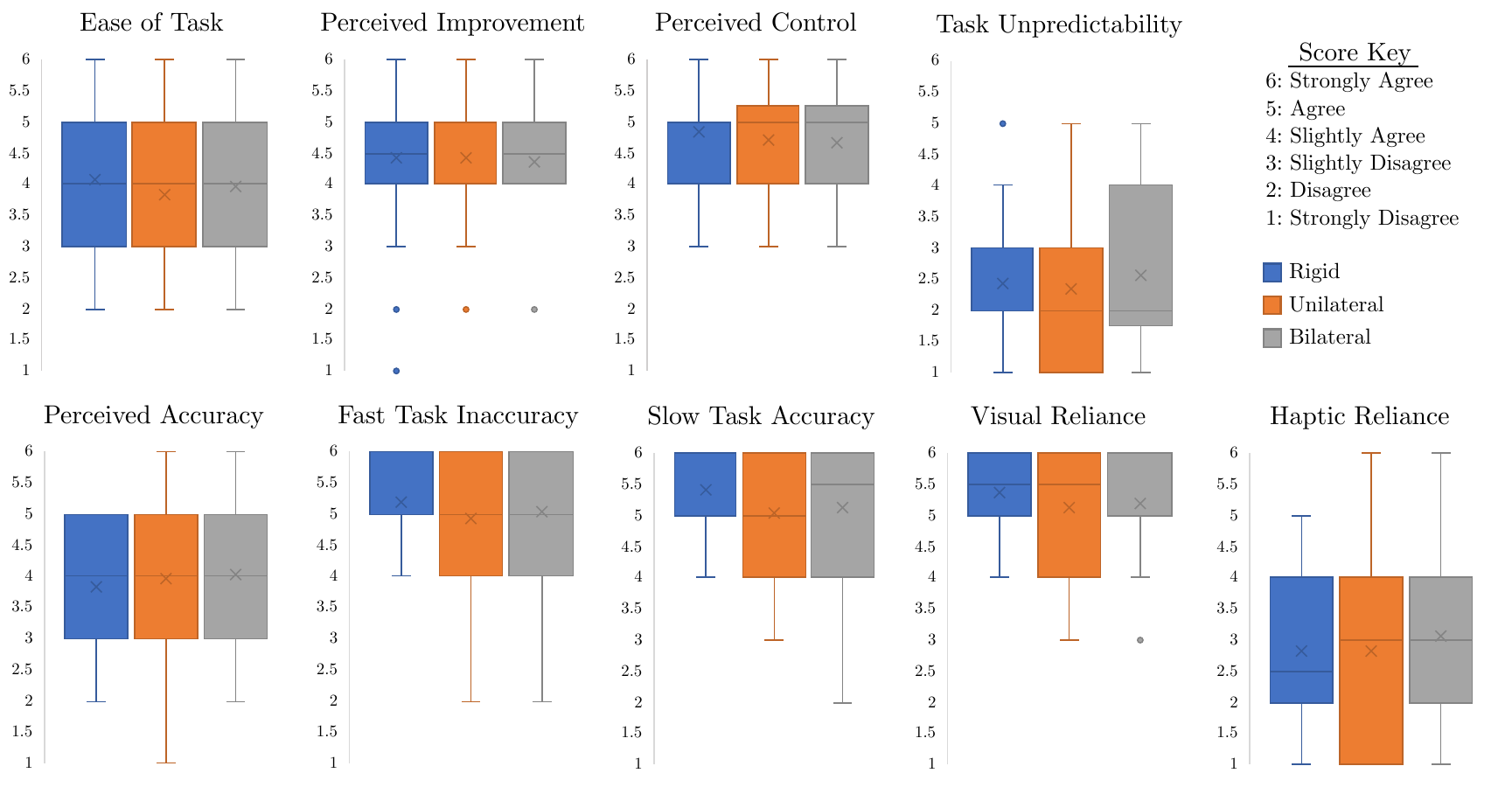}
   \caption{Agreement scores from users in post-task survey, scaled from 1 (Strongly Disagree) to 6 (Strongly Agree) and clustered into transmission mode by color. Scores are aggregated across individuals, with the mid-box lines as sample medians, and vertical x-mark positions indicating sample means.}
   \label{fig:teleopsurvey}
\end{figure*}

Box-and-whisker charts of response distributions from the post-configuration surveys is shown in Figure~\ref{fig:teleopsurvey}. While it appears dynamics do not play much a role in multiple aspects of perceived ability and task interaction, users slightly agreed more that Rigid transmission helped their task accuracy compared to other EM modes. Also, users did not report much reliance on haptic (kinesthetic) information during the task, with scores centered around ``Slightly Disagree''.

\section{Discussion}
Our addition of different oscillation speed profiles as a proxy for task difficulty did test the human limits of teleoperative object tracking. As the oscillation sped up or increased in component complexity, the user altered their exerted pinch grip as they modified their grasp strategy to maintain performance. The gain error from ideal target tracking at both teleoperator ports increased as oscillation became more difficult, and the synchronization represented by phase error worsened at especially higher frequencies above\,1.75 Hz. Thus, our third prediction stands since task difficulty from target speed did impact performance. These results of performance breakdown over task speed generally agree with hand joystick tracking observations in prior literature \cite{Pew1974HumanPerformance}. Whereas our original study \cite{Singhala2023TelerobotTask} only tested a composite trajectory of 3 components from the low end of frequencies tested by previous hand joystick tracking \cite{Pew1974HumanPerformance}, our new study filled a larger gamut of task frequency speeds up to when tracking was subjectively too fast. This study's users also subjectively reported that they are inaccurate during fast tracking and accurate during slow tracking, so there was a conscious component to this decreasing performance.

While the influence of configuration on tracking gain and phase error between the environment-facing follower shaft and the virtual target cannot be conclusively confirmed (but still assumed to be approximately the same) from this analysis, we did see statistically significant performance changes in matching gain between the user-facing leader port and target between transmission modes. Conditionally, our first prediction does prove true for leader/user adjustment performance between configurations of dynamics. These findings on follower and leader performance generally agree with our team’s previous tracking study on time-domain error \cite{Singhala2023TelerobotTask}. This highly suggests that the user was adjusting their tracking strategy between transmissions, specifically direct rigid connection and unilateral EM transmission without kinesthetic feedback. Together with changed gain performance with more task repetitions, this implies some user incorporation or compensation of the teleoperative tool as familiarity increased over time. Post-configuration responses on positive perceived improvement across all transmissions support this conclusion. Broadly speaking, EM dynamics introduced delays and varying amounts of position or torque scaling between the operator and the virtual disk. Therefore, in order to accurately perform the tracking task, the operator had to compensate for these dynamics as they rotated the hand fixture. Further supplementing this argument of compensating dynamics is the fact that our participants used different grip forces when using the unilateral (no force-reflection) electromechanical teleoperator as compared to the bilateral configuration, demonstrating our second prediction about grip differences between configurations to be correct. This observation supports the theory that humans match their limb impedance to the impedance of their tool, resulting in changes in grip effort. In future work, we can use sEMG equipment in addition to end-effector squeeze sensing to capture a stronger state estimate of arm impedence.

Our quantitative task performance findings are also substantiated by our qualitative findings. In particular, participants reported a higher reliance on visual feedback to complete the task. While visual dominance is expected in a visuo-haptic tracking task, it seems as though participants used haptic feedback much less than expected; on average, participants disagreed with the statement that they found haptic feedback to be useful for either of the two Electromechanical teleoperators. (This observation agrees with surveys in our previous study too \cite{Singhala2023TelerobotTask}.) It is therefore possible that participants were learning to invert or compensate for the model of the teleoperator (and incorporate the teleoperator’s dynamics) based more on visual observation more than kinesthetic feedback, but both sensations still play a role. Alternatively, laypersons may understand “haptic” as vibrational rather than kinesthetic, so more careful language should be used in the future. Users also generally slightly agreed that the tool is easy to use and control between transmissions, which may imply that the process of tool incorporation and compensation does not take much physical or cognitive effort. Therefore, the process of incorporating different dynamics was intuitive for users.

From statistical significance of fixed-effect coefficients of our linear models with interaction terms between configuration and frequency, we can start to reveal clues about how task dynamics and speed interact. Given that grip force and gain error for both ports and phase error at the leader port had statistically significant covariate differences in model coefficients, this may imply that performance changes across different dynamics occur at specific frequencies of interest, whereas aggregating across all difficulty speeds ends up being inconclusive. This might reveal that different performance drop curves as a function of tracking speed qualitatively seen in Figures~\ref{fig:teleopFollowSiS}a and \ref{fig:teleopLeaderSiS}a may actually be statistically significant for an entire healthy human population.

By testing the limits of human tracking in this experiment, there is promising work in isolate measures of proprioceptive ability from general sensorimotor ability. It is possible to use multiple transmission connections in parallel during teleoperation. Specifically, you can normally use a rigid connection to transmit torque between the user and the environment, but then inject sudden torque disturbances with connected motors and observe how users try to recover their intended motion. A lower-limb analogy would be walking a treadmill with a comfortable gait, but then experiencing a yanking impulse at the waist and having to try to recover balance.

One recent study that investigated predictive feedforward control in reference tracking found that the tracking control from human users approximately followed superposition of disturbance and reference input \cite{Yamagami2023EffectTrajectory-Tracking}. A limitation of our current study is that we did not directly control either visual or kinesthetic disturbance (i.e. the human user was not given intentionally distorted feedback). By controlling disturbance in future studies with more model training and prediction trials per condition, it should be easier to generate more accurate models of combined feedback-feedforward control and to verify linear time-invariant (LTI) transformations during wrist pursuit teleoperation. Furthermore, we could run this study on persons affected by cerebellar ataxia, a condition known to affect predictive control of motion. By observing this group’s sensorimotor response, it is possible to infer the importance that feedforward control has in a teleoperative human-in-the-loop system. From observing the behavior of users with neurological disease more generally, it is possible to see how neurological pathway alteration affects the ability to incorporate tools and apply agency.

{\appendices
\section*{Appendix A: Follower-Target Response by Configuration}
\label{appendix:followcfg}
\begin{table}[H]
  \centering
  \caption{P-values from pairwise comparisons of magnitude tracking at follower port between configurations.}
    \begin{tabular}{lrr}
    \toprule
    Cfg & \multicolumn{1}{r}{Rigid} & \multicolumn{1}{r}{Unilateral} \\
    \midrule
    Bilateral & 0.060 & 0.999 \\
    Rigid &       & 0.056 \\
    \bottomrule
    \end{tabular}%
  \label{tab:stkabscfg}%
\end{table}%

\begin{table}[H]
  \centering
  \caption{P-values from pairwise comparisons of tracking phase error at follower port between configurations.}
    \begin{tabular}{lrr}
    \toprule
    Cfg & \multicolumn{1}{r}{Rigid} & \multicolumn{1}{r}{Unilateral} \\
    \midrule
    Bilateral & 0.477 & 0.963 \\
    Rigid &       & 0.330 \\
    \bottomrule
    \end{tabular}%
  \label{tab:stkphscfg}%
\end{table}%

\section*{Appendix B: Leader-Target Response by Configuration}
\label{appendix:leadercfg}
\begin{table}[H]
  \centering
  \caption{P-values from pairwise comparisons of tracking phase error at leader port between configurations.}
    \begin{tabular}{lrr}
    \toprule
    Cfg & \multicolumn{1}{r}{Rigid} & \multicolumn{1}{r}{Unilateral} \\
    \midrule
    Bilateral & 0.995 & 1.000 \\
    Rigid &       & 0.993 \\
    \bottomrule
    \end{tabular}%
  \label{tab:ldrphscfg}%
\end{table}%

\section*{Appendix C: Phase Responses by Repeated Task Trials}
\label{appendix:phstr}
\begin{table}[H]
  \centering
  \caption{P-values from pairwise comparisons of tracking phase error at follower port between trials.}
    \begin{tabular}{lrrr}
    \toprule
    \multicolumn{1}{l}{Trial} & 2     & 3     & 4 \\
    \midrule
    1     & 0.887 & 0.085 & 0.158 \\
    2     &       & 0.337 & 0.499 \\
    3     &       &       & 0.993 \\
    \bottomrule
    \end{tabular}%
  \label{tab:stkphstrial}%
\end{table}%

\begin{table}[H]
  \centering
  \caption{P-values from pairwise comparisons of tracking phase error at leader port between trials.}
    \begin{tabular}{lrrr}
    \toprule
    \multicolumn{1}{l}{Trial} & 2     & 3     & 4 \\
    \midrule
    1     & 0.743 & 0.204 & 0.423 \\
    2     &       & 0.747 & 0.941 \\
    3     &       &       & 0.974 \\
    \bottomrule
    \end{tabular}%
  \label{tab:ldrphstrial}%
\end{table}%

\begin{figure}[H] 
    \centering
    \includegraphics[scale=0.3]{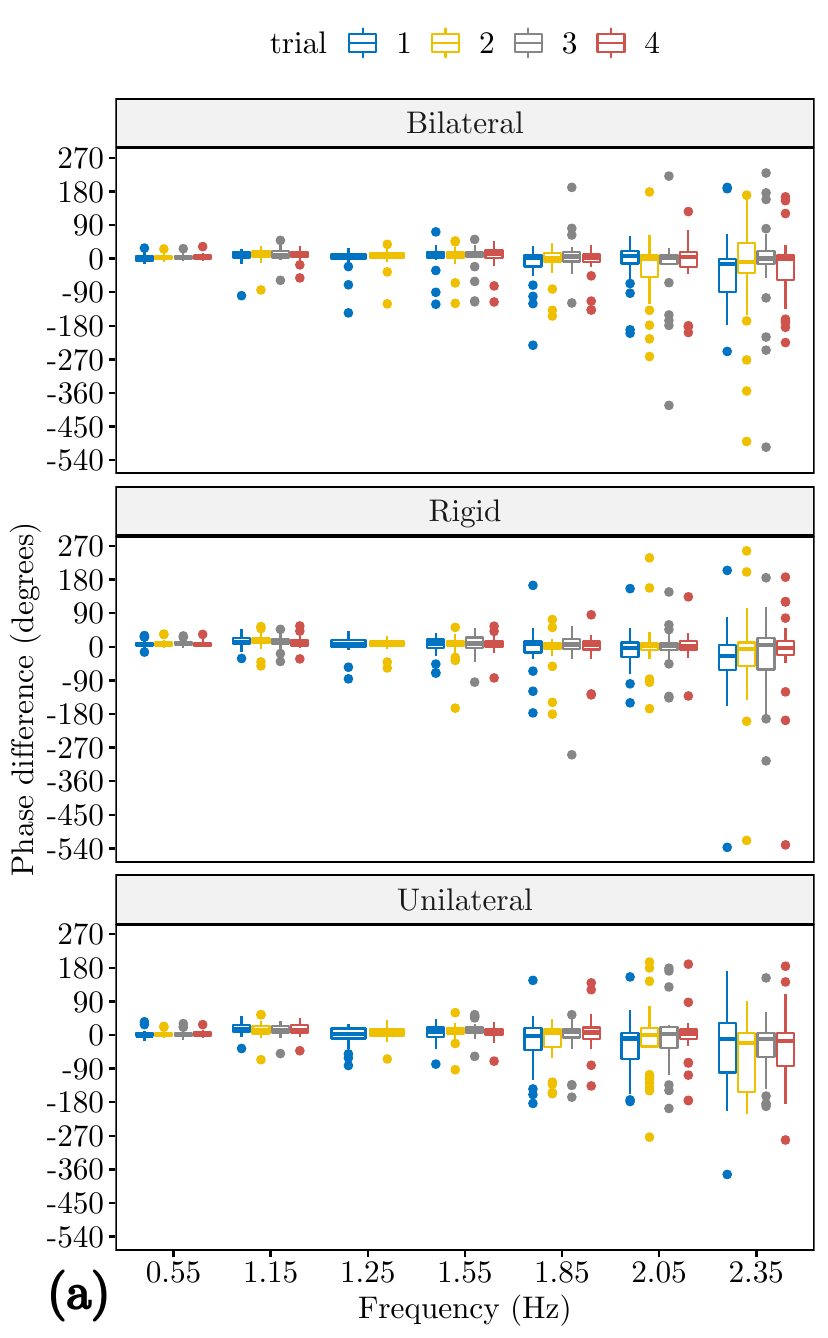}
    \includegraphics[scale=0.3]{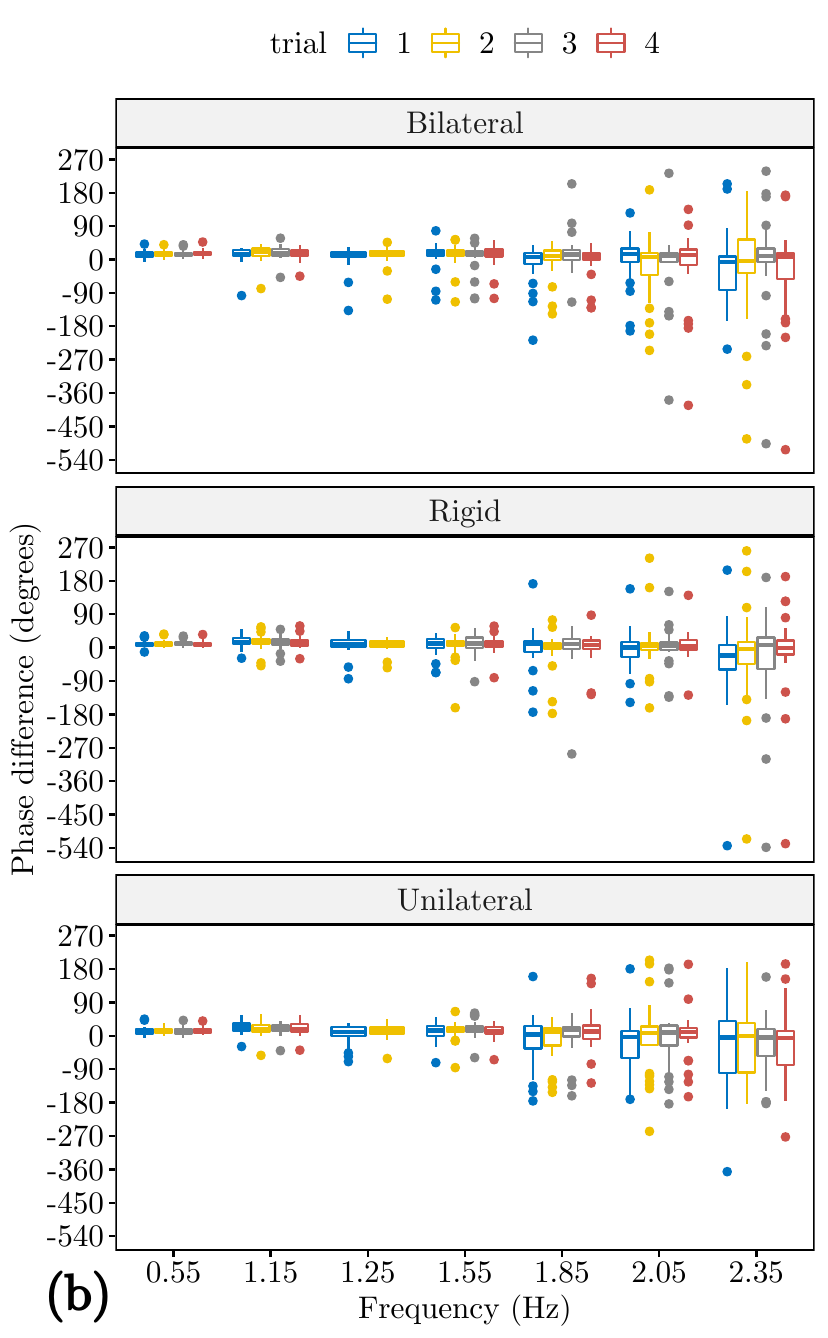}
    \caption{Modified Bode plots of the frequency responses expressing relative phase error for (a) follower-target transfer function and (b) leader-target transfer function. Responses for all three transmission configurations are indicated by facet. Error ratios are aggregated across participants, but clustered by repeated task count into separate colors.}
    \label{fig:teleopSiSphstr}
\end{figure}
}

\bibliographystyle{IEEEtran}
\bibliography{references.bib}

\end{document}